\documentclass[hidelinks,onefignum,onetabnum]{siamart251216}
\usepackage[paperwidth=7in,left=2cm, right=2cm, top=3cm, bottom=1cm]{geometry}

\usepackage{amsmath}
\usepackage{amsfonts,amssymb}
\usepackage{booktabs}       
\usepackage{multirow}
\usepackage{makecell}
\usepackage{algpseudocode}
\usepackage{wrapfig}

\makeatletter
\let\@fnsymbol\@arabic

\newtheorem{remark}{Remark}

\title{
A Bifurcation-Based Domain Decomposition Method with Neural Operators for Blood Flow Simulation 
}
\author{
Yuzhou Zhao \thanks{Department of Electrical and Computer Engineering, Princeton University, NJ, USA.
(\email{yz4450@princeton.edu})}
\and
Han Zhang \thanks{Corresponding Author. Division of Applied Mathematics, Brown University, Providence, RI, USA.
(\email{han\_zhang6@brown.edu})}
\and
J. Matías Di Martino \thanks{Department of Computer Science, Universidad Cat\'olica del Uruguay, Montevideo, Uruguay, and Department of Electrical and Computer Engineering, Duke University, NC, USA.
(\email{matias.di.martino@duke.edu})}
\and
Jean-Michel Morel \thanks{Division of Industrial Data Science, Lingnan University, Hong Kong, China.
(\email{jeanmichelmorel@ln.edu.hk})}
\and
Guillermo Sapiro \thanks{Department of Electrical and Computer Engineering and Princeton Precision Health, Princeton University, NJ, USA; and Apple, CA, USA.
(\email{guillermos@princeton.edu})}
}
\begin{document}
\maketitle
\begin{abstract}
Fast and accurate simulation of hemodynamic behavior within vascular networks is essential for numerous clinical applications. However, obtaining high-quality and computationally efficient flow measurements across complex vascular networks remains challenging. To address this, we first decompose the vascular network into a set of bifurcation units and then develop an operator network capable of mapping unit-specific parameters to the local solution fields of each bifurcation unit. By lumping the Windkessel-model outlet parameters and incorporating inlet boundary conditions from the solution of parent units, the flow and pressure fields can be rapidly approximated. Subsequently, operator-network-driven Schwarz waveform relaxation is applied across bifurcation units to correct discontinuities and improve numerical accuracy. On 7-segment and 55-segment arterial tree models, the proposed method achieves $13\times$ to $17\times$ wall-clock speedups over conventional 1D numerical simulation, with relative $L^2$ errors of 1\% in both pressure and velocity. The resulting pulse wave velocity biomarkers agree with the conventional reference to within 1--2\%, and the same trained operator 
generalizes to different tree-like 1D vascular network topologies.
\end{abstract}

\section{Introduction}

Blood flow simulation plays a fundamental role in characterizing cardiovascular dynamics, providing critical insights into hemodynamic behavior within the vascular system. It is indispensable for example for medical monitoring, cardiovascular disease diagnosis, and surgical planning. In recent years, a wide range of data-driven and machine learning methods have been developed to accelerate blood flow simulation and enhance its predictive capabilities~\cite{pegolotti2024learning,velikorodny2025deep}, offering new opportunities for efficient and personalized cardiovascular analysis.

However, existing supervised learning approaches rely heavily on large amounts of high-fidelity training data, which are expensive and time-consuming to generate. In particular, obtaining accurate simulation data for a single complex vascular network often requires hours to days of computational effort. Moreover, these data-driven models typically lack robustness with respect to geometric variations; when the configuration or topology of the target vascular network changes, the model must be retrained or significantly adapted. This limitation hinders their applicability in realistic scenarios, where the geometries of interest vary widely across subjects, calling for more flexible and generalizable modeling frameworks.

Since obtaining the solution field over a complete vascular network is computationally challenging, in this work we adopt a modular perspective and focus on learning the behavior of recurring structural patterns within the network. In particular, we identify common local structures (motifs or units) that frequently arise in vascular systems, namely bifurcations. For these decomposed units, we construct an operator-based model that learns the mapping from unit-specific parameters to the local solution field. By treating these common structures as reusable building blocks, the global solution over a complex vascular network can be approximated by assembling the solutions predicted on each unit, which follows the concept of the domain decomposition approach~\cite{chan1994domain}. This strategy avoids the need to learn a global model for every network configuration and provides a scalable pathway toward efficient blood flow simulation.

In this work, we adopt a reduced-order modeling approach based on the 1D Navier-Stokes equations~\cite{quarteroni2004mathematical}, which provides an efficient and practical framework for capturing blood flow dynamics in vascular networks. In its dimensionless form, the governing model depends on a set of parameters that include physical properties (e.g., density and viscosity), geometric characteristics (e.g., vessel length and reference cross-sectional area), and boundary conditions prescribed at the branch inlet and outlet.

\begin{wrapfigure}{r}{0.6\textwidth}
    \centering
    \vspace{-1.5em}
    \includegraphics[width=0.6\textwidth]{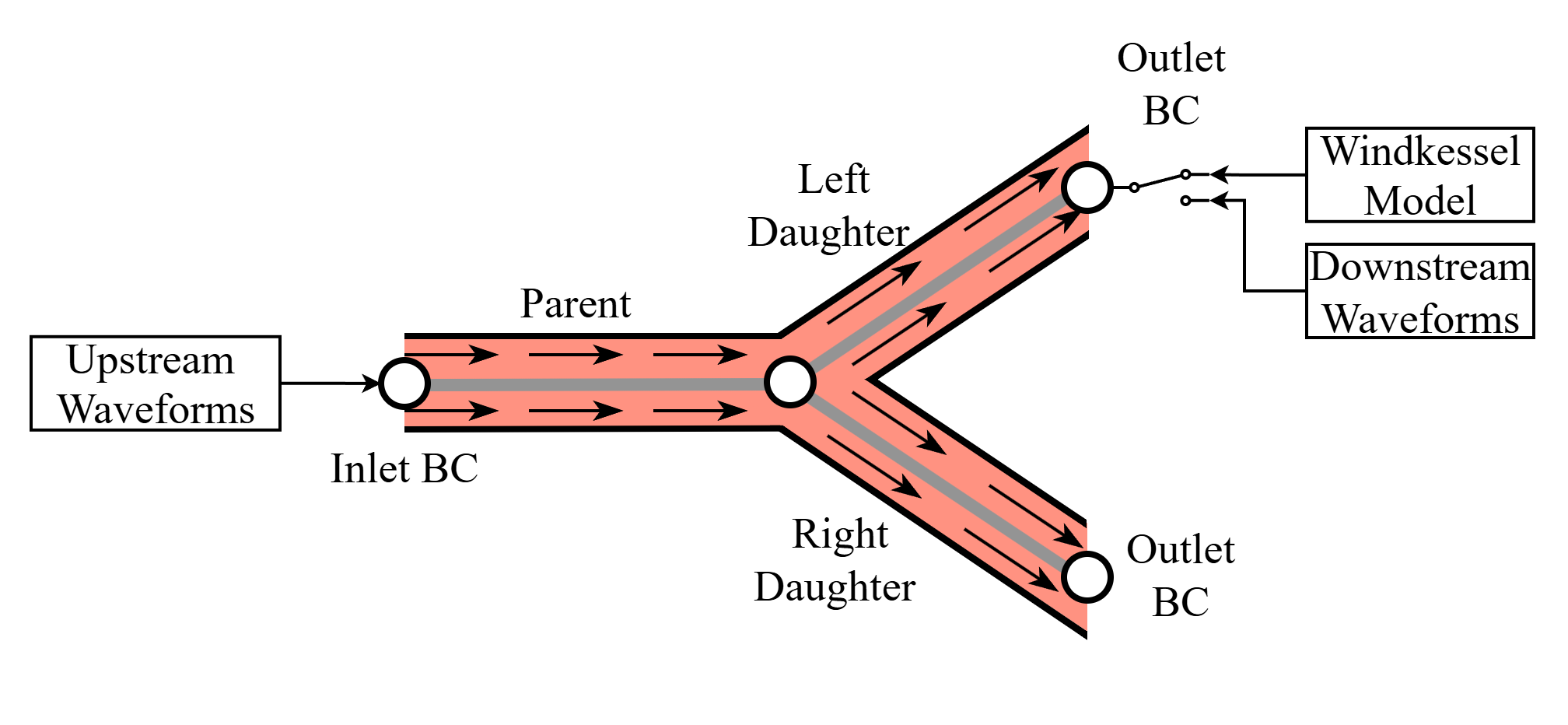}
    \vspace{-2.2em}
    \caption{A single bifurcation unit (BU). The parent segment receives an inlet boundary condition from upstream waveforms and bifurcates into two daughter segments. Each outlet is equipped with either a Windkessel (RCR) model or downstream waveforms from a child BU.}
    \vspace{-0.5em}
    \label{fig:bu_schematic}
\end{wrapfigure}

Within this setting, we adopt bifurcation units (BUs) (Fig.~\ref{fig:bu_schematic}) as the fundamental building blocks for domain decomposition. A key motivation is to avoid explicitly deriving flow-splitting relations at junctions, which would otherwise be required when treating individual vessel segments as units. In this setup, the inlet boundary conditions for a child unit can be directly obtained from its parent, as the corresponding interface lies entirely within the parent domain. For outlet conditions, we employ Windkessel models to represent downstream vascular effects, and parameter lumping systematically aggregates distal parameters into equivalent upstream values. By recursively propagating this procedure from terminal (leaf) nodes toward the root, we obtain consistent and effective boundary conditions for each unit.

Our goal is to train an operator network capable of mapping unit-specific parameters to the local solution field for different BU. The overall algorithm proceeds in three main stages. First, the Windkessel outlet parameters for each unit are lumped together with geometric, physical, and boundary parameters to form a consistent set of unit-specific inputs. Second, each BU is solved individually (potentially in parallel) using the trained operator network, which rapidly approximates the flow and pressure fields without requiring high-fidelity numerical simulations. Third, Schwarz waveform relaxation (SWR) is performed across BUs to correct any inconsistencies or discontinuities in the overlapping regions. Notably, all computations during these iterations are also carried out using the operator network, ensuring computational efficiency while maintaining accuracy.

We validate the proposed framework on two vascular models, namely a 7-segment arterial network and the 55-segment full arterial tree of \cite{Willemet2015}. The operator-driven SWR solver achieves $13\times$ to $17\times$ wall-clock speedups over conventional 1D numerical simulation, with mean relative $L^2$ errors of the order of 1\% in both pressure and velocity. Our empirical experiments show that, within the settings considered in this work, multiplicative SWR converges for the 1D Navier--Stokes blood flow equations on vascular networks. Tests also show that the locality of the BU operator design enables generalization to incomplete vascular topologies without retraining. The impact of this proposed computational framework on downstream tasks such as the computation of pulse wave velocity, a biomarker of arterial health, is also investigated.

To summarize, our contributions are as follows:
\begin{itemize}
    \item We propose a novel strategy to decompose the vascular network into BUs, establishing a modular structure that facilitates efficient computation.
    \item We formulate an SWR algorithm with lumped Windkessel parameters for the decomposed BUs. Our empirical experiments show that, on vascular networks under our setting, multiplicative SWR converges for the 1D Navier--Stokes blood flow equations.
    \item We develop a BU operator network capable of predicting the local solution fields for each BU, replacing the numerical solver with a single forward pass per unit.
    \item We demonstrate $13\times$ to $17\times$ speedups over conventional 1D numerical simulation, while maintaining downstream biomarkers consistent with the numerical reference solutions. We further show that the method generalizes to incomplete vascular topologies.
\end{itemize}
\section{Related Work}

\subsection{Blood Simulation}
Accurate blood flow measurement is vital for cardiovascular healthcare, informing diagnosis, surgical planning, and treatment. Although invasive methods such as Fractional Flow Reserve (FFR) \cite{pijls1996measurement} are widely used, they pose risks such as medical complications and can disrupt the dynamics of natural flow \cite{yan2022impact}, leading to measurement inaccuracies. Numerical simulations offer a noninvasive alternative by modeling blood as an incompressible Newtonian fluid governed by the Navier-Stokes equations \cite{peskin1972flow}. 

To account for vessel deformation, these models are extended to fluid-structure interaction (FSI) frameworks \cite{quarteroni2004mathematical}. Methods have been proposed to solve FSI problems. These include dimensional coupling \cite{formaggia2001coupling}, parallel preconditioning \cite{crosetto2011parallel}, and splitting schemes with stability guarantees \cite{bukavc2013fluid}. Other approaches involve space-time formulations \cite{bazilevs2008isogeometric}, discontinuous Galerkin methods \cite{wang2018higher}, coupled momentum methods \cite{figueroa2006coupled}, and learning-based approaches \cite{zhang2024meshless,zhang2024learning}. 

Reduced-order models based on the 1D Navier-Stokes equations have been widely developed for efficient hemodynamic simulations \cite{formaggia2002one,formaggia2003one}. The physical consistency of such 1D models with respect to their 3D counterparts has been systematically investigated \cite{formaggia2013physical}. In addition, comparisons with full 3D simulations incorporating Windkessel models to represent outlet boundary conditions have been conducted \cite{grinberg2011modeling}. More recently, similar analysis has been carried out in the context of physics-informed neural network approaches \cite{zhang2025parametrized}. In \cite{perdikaris2014fractional}, fractional-order viscoelastic formulations have been introduced to enhance the description of vascular wall behavior, where it has been shown that the hemodynamic response is strongly influenced by the fractional order parameter. Multi-scale modeling strategies have also been proposed, coupling 3D and 1D formulations to significantly reduce computational cost while retaining essential physical fidelity \cite{formaggia2001coupling}. Furthermore, data-driven approaches such as graph neural networks have been employed to predict pressure and flow evolution at nodal points based on local features and current states \cite{pegolotti2024learning}. The computational efficiency of 1D models has enabled a range of downstream applications that require large ensembles of forward simulations, including uncertainty quantification \cite{fleeter2020multilevel,brault2017uncertainty}, virtual patient cohort generation \cite{Willemet2015,charlton2019modeling}, physics-informed parameter estimation from clinical measurements \cite{kissas2020machine}, and noninvasive fractional flow reserve prediction \cite{fossan2021machine}.

Despite this progress, existing methods, including graph neural network-based approaches, either treat the entire vascular network as a monolithic system—limiting scalability to unseen topologies—or predict only nodal quantities without resolving full spatiotemporal solution fields within individual vessels. In contrast, our work decomposes the vascular network into reusable bifurcation units and learns operators that map unit-specific parameters to local pressure and velocity fields, enabling topology-generalizable simulation for tree-like 1D vascular networks without retraining.

\subsection{Operator Learning}

Operator learning methods \cite{goswami2023physics,lu2021learning} aim to generalize across families of partial differential equations by directly approximating mappings between function spaces. Representative examples include DeepONet \cite{lu2021learning}, which employs a branch–trunk architecture, and the Fourier Neural Operator \cite{li2020fourier}, which efficiently captures long-range dependencies in Fourier space. From the idea of Physics-Informed Neural Networks \cite{raissi2019physics}, \cite{wang2021learning} introduces a physics-informed DeepONet by constraining the network outputs to satisfy the underlying governing equations. More recently, \cite{wang2024cvit} proposes a continuous vision transformer that leverages advances in computer vision to improve the modeling of physical systems.

In terms of applications, \cite{cai2021deepm} proposes a data assimilation framework for solving multiphysics and multiscale electroconvection problems, achieving significantly faster predictions compared to conventional numerical methods through the use of pre-trained neural networks. The work in \cite{koric2023data} applies DeepONet to solve the heat conduction equation with parametrized, spatially varying multidimensional heat sources. In \cite{cheng2025surrogate}, a surrogate model is developed to predict heat transfer under fluctuating flow conditions. Furthermore, \cite{rezaei2025digitalizing} employs physics-informed operator learning to map digital microstructures to mechanical fields, enabling rapid prediction of microstructural responses based on realistic material topologies. Lastly, \cite{abueidda2025deepokan} extends the DeepONet framework to model stress distributions in small-strain plasticity problems, incorporating variable loads, material properties, and stochastic deformation paths as input parameters.

These operator learning methods have been applied to single-domain or fixed-geometry problems, where the computational domain does not change across evaluations. None of the above approaches address the setting of a domain-decomposed hemodynamic system in which the operator must generalize across bifurcation units with varying geometry, lumped boundary conditions, and network topology. Our work fills this gap by designing a bifurcation-unit operator that maps unit-specific parameters to local solution fields and can be composed via Schwarz iterations to solve arbitrary vascular networks.

\subsection{Domain Decomposition Methods and Schwarz Waveform Relaxation}

Domain decomposition methods (DDM) solve large-scale PDEs by partitioning the computational domain into overlapping or non-overlapping subdomains, solving local subproblems independently, and enforcing consistency at the interfaces through iterative exchange of boundary data~\cite{chan1994domain,quarteroni1999domain}. The classical Schwarz method, in both its additive and multiplicative variants, operates on steady-state or time-stepping formulations, where at each iteration the subdomain solutions are updated using the most recent interface values from neighboring subdomains.

Schwarz waveform relaxation (SWR) extends the classical Schwarz framework to time-dependent problems by exchanging entire space-time traces (waveforms) at subdomain interfaces, rather than pointwise-in-time boundary values~\cite{gander1998space,gander2003optimal,gander2007optimized}. Each subdomain solves its local initial-boundary value problem over the full time window, and the interface waveforms are updated iteratively until convergence. This formulation naturally accommodates different time discretizations or solvers on each subdomain and has been theoretically analyzed for the heat equation~\cite{gander1998space}, wave equations~\cite{gander2003optimal}, and advection-reaction-diffusion systems~\cite{gander2007optimized,caetano2010schwarz,descombes2011schwarz}.

SWR has been applied to coupled multi-physics systems, including partitioned surface-coupled multiphysics applications~\cite{rueth2021quasi}, coupled advection-diffusion-reaction equations~\cite{dong2024convergence}, and the primitive equations of the ocean~\cite{lemarie2010optimized}. However, to our knowledge, it has not been applied to the 1D Navier-Stokes blood flow equations on vascular networks prior to this work. Moreover, in all the above applications, the subdomain solvers remain conventional numerical methods, while we apply a surrogate model to solve the decomposed units. 

\section{Problem and Preliminaries}

\begin{table*}[ht!]
\footnotesize
\centering
\vspace{-2.0pc}
\caption{Summary of notations.}
\begin{tabular}{p{0.09\textwidth}p{0.34\textwidth}|p{0.09\textwidth}p{0.34\textwidth}}
\hline
\textbf{Symbol} & \textbf{Description}          & \textbf{Symbol}   & \textbf{Description} \\
\hline
$\mathcal{V}$   & Set of nodes                  & $\mathcal{L}$     & Set of vessel segments \\
$v_i$           & Node indexed by $i$           & $\Omega_i$        & Vessel segment end with $v_i$ \\
$L_i$           & Length of segment $\Omega_i$  & $A_i$             & Cross-sectional area on $\Omega_i$ \\
$Q_i$           & Flux rate on $\Omega_i$       & $P_i$             & Pressure on $\Omega_i$ \\
$\rho_i$        & Blood density                 & $\mu_i$           & Dynamic viscosity\\
$\alpha_i$      & Momentum correction coefficient& $K_i$           & Viscous resistance parameter \\
$A_{d,i}$       & Reference (diastolic) area    & $P_{d,i}$         & Diastolic pressure \\
$\beta_i$       & Wall stiffness parameter      & $E_i$             & Young's modulus \\
$h_i$           & Vessel wall thickness         & $c_i$             & Pulse wave velocity (PWV) \\
$R_{1,i}$       & Proximal resistance           & $R_{2,i}$         & Distal resistance\\
$C_i$           & Compliance                    & $Z_i$             & Characteristic impedance \\
$R_{T,i}$       & Total resistance              & $C_{T,i}$         & Total compliance\\
$R_{v,i}$       & Vessel resistance contribution& $C_{v,i}$        & Vessel compliance contribution \\
\hline
\end{tabular}
\label{tab:notations}
\end{table*}

\begin{figure}
    \centering
    \includegraphics[width=0.78\linewidth]{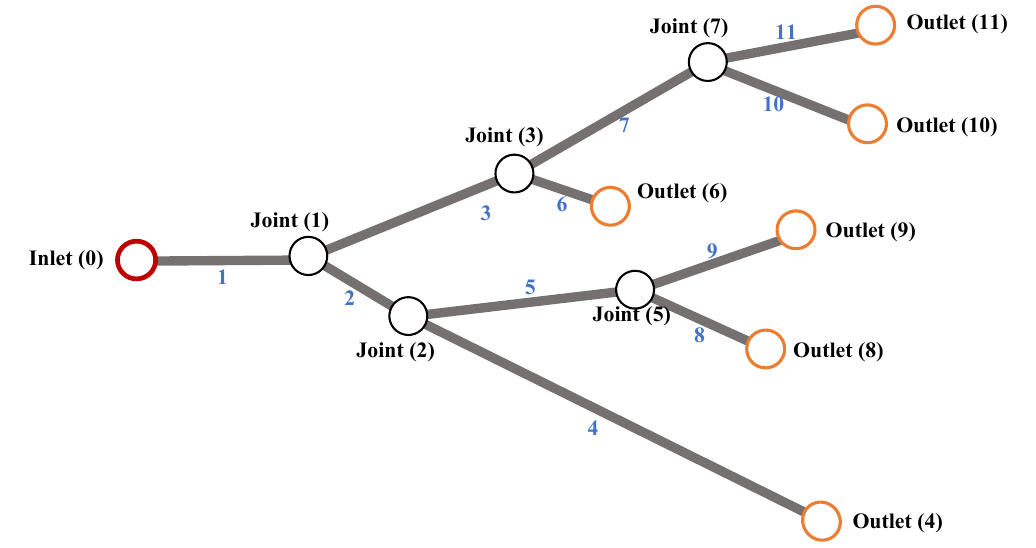}
    \caption{Illustration of a simplified vessel tree.}
    \label{fig:vessel_tree}
\end{figure}

Table~\ref{tab:notations} summarizes the notations used throughout this manuscript. We consider a vascular network modeled as a collection of one-dimensional compliant vessel segments connected through junctions. The network is represented by a set of nodes $\mathcal{V} = \{v_i\}_{i=0}^{N}$, which include inlet nodes, junction nodes, and outlet nodes (Fig.~\ref{fig:vessel_tree}). To describe the vessel segments, we introduce the set $\mathcal{L} = \{\Omega_i\}_{i=1}^{N}$, where each segment $\Omega_i$ is uniquely associated with its outlet node $v_i$ and has length $L_i$, defined over the domain $[0, L_i]$. This indexing convention ensures a one-to-one correspondence between segments and indices, as a node may serve as the inlet for multiple segments but, under a directed tree structure, can be the outlet of at most one segment, with the exception of the root node $v_0$. 

For convenience of presentation, we further partition the node set into three disjoint subsets as $\mathcal{V} = \{v_0\} \cup \mathcal{V}^{\mathrm{conj}} \cup \mathcal{V}^{\mathrm{out}}$, where $\mathcal{V}^{\mathrm{conj}}$ denotes the set of conjunction nodes, and $\mathcal{V}^{\mathrm{out}}$ denotes the set of terminal (leaf) nodes corresponding to vessel outlets. We denote by $M = |\mathcal{V}^{\mathrm{conj}}|$ the number of conjunction nodes.

Specifically, we restrict our attention to vascular networks that can be represented as binary trees, where each junction connects one parent vessel segment and two child segments. This assumption is made without loss of generality, as more complex junctions can be equivalently reformulated within this framework. In particular, a multi-branch junction (e.g., a trifurcation) can be decomposed into a sequence of bifurcations by introducing auxiliary segments, where one of the intermediate segments may have negligible (or zero) length. This representation allows us to treat tree-like 1D vascular topologies in a unified manner while maintaining a consistent bifurcation-based formulation for analysis and computation.

\subsection{Blood Flow via 1D Navier-Stokes}

Under the standard assumptions of an impermeable vessel wall and an incompressible Newtonian fluid, the conservation of mass and momentum applied to a control volume yields the one-dimensional Navier-Stokes equations. The primary unknowns are the cross-sectional area $A$, volumetric flow rate $Q$, and pressure $P$. On each vessel segment $\Omega_i$, we denote the restricted variables by $A_i(x,t) = A|_{\Omega_i}(x,t)$, $Q_i(x,t) = Q|_{\Omega_i}(x,t)$, and $P_i(x,t) = P|_{\Omega_i}(x,t)$. The governing equations on each segment $\Omega_i$ are given by
\begin{equation}
\begin{aligned}
\frac{\partial A_i}{\partial t} + \frac{\partial Q_i}{\partial x} &= 0, 
\\
\frac{\partial Q_i}{\partial t} + \frac{\partial}{\partial x}\!\left(\alpha_i \frac{Q_i^2}{A_i}\right) 
+ \frac{A_i}{\rho_i}\frac{\partial P_i}{\partial x} 
&= - K_i \frac{Q_i}{A_i},
\end{aligned}
\label{eq:1dns_segment}
\end{equation}
where $\rho_i$ denotes the blood density, $\alpha_i$ is the momentum correction coefficient, and $K_i = 8\pi\mu_i/\rho_i$ is the viscous resistance parameter, with $\mu_i$ being the dynamic viscosity.

\noindent Assuming the arterial wall to be a linear elastic, thin, incompressible, and homogeneous solid, we obtain the tube law describing a pressure-area relationship for $\Omega_i$,
\begin{equation}
P_i = P_{\mathrm{d},i} + \frac{\beta_i}{A_{d,i}}(\sqrt{A_i} - \sqrt{A_{d,i}}),
\label{eq:wall}
\end{equation}
where $A_{d,i}$ is the reference area under diastolic pressure $P_{\mathrm{d},i}$, and $\beta_i$ denotes the stiffness of segment $\Omega_i$ with
\begin{equation}
    \beta_i = \frac{4}{3} \sqrt{\pi} E_i h_i,
\end{equation}
where $E_i$ denotes the elastic modulus and $h_i$ the thickness of the vessel wall.

\noindent The inlet boundary condition should be some given flux historic profile given by prior measurements, which is given as a Dirichlet condition
\begin{equation}
Q_1(0,t) = Q_{in}.
\label{eq:inlet}
\end{equation}

\noindent The initial condition is set to start from a quiescent reference state $P_{d,i}$ and $Q_{d,i}$, and is given for each $\Omega_i$ by
\begin{equation}
P_i(x,0) = P_{d,i}, \qquad Q_i(x,0) = Q_{d,i}.
\end{equation}

\noindent At each junction node $v_i$ (i.e., neither an inlet nor an outlet) connecting the segment $\Omega_i$ with its two children segments $\Omega_{i_1}$ and $\Omega_{i_2}$, where $i_1$ and $i_2$ denote the indices of the children vessels of vessel $i$, we enforce conservation of mass and continuity of pressure. The mass conservation condition requires that the flow in the parent vessel equals the sum of the flows in the daughter vessels. The pressure continuity condition requires that the pressures at the junction are equal across all connected segments. The coupling conditions at the junction can be written as
\begin{equation}
Q_{i}(L_i,t) = Q_{i_1}(0,t) + Q_{i_2}(0,t), \qquad 
P_{i}(L_i,t) = P_{i_1}(0,t) = P_{i_2}(0,t).
\label{eq:conjunction}
\end{equation}

\noindent For vessel segments $\Omega_i$ terminating at the outlets, where $v_i \in \mathcal{V}^{\mathrm{out}}$, we model the outlet boundary conditions using a three-element Windkessel (RCR) model \cite{westerhof1971artificial,westerhof2009arterial}, consisting of a proximal resistance $R_{1,i}$, a distal resistance $R_{2,i}$, and a compliance $C_i$. The corresponding boundary condition is given by
\begin{equation}
\left(1 + \frac{R_{1,i}}{R_{2,i}}\right) Q_{i}
+ R_{1,i} C_i \frac{\partial Q_{i}}{\partial t}
=
\frac{P_{i} - P_\text{inf}}{R_{2,i}}
+ C_i \frac{\partial P_{i}}{\partial t}.
\label{eq:outlet}
\end{equation}
Where $P_\text{inf}$ denotes the distal reference pressure of the downstream vascular bed, representing the asymptotic pressure toward which the system relaxes.

\subsection{Parameter Lumping}
\label{sec:rcr_lumping}
\begin{figure}
    \centering
    \includegraphics[width=\linewidth]{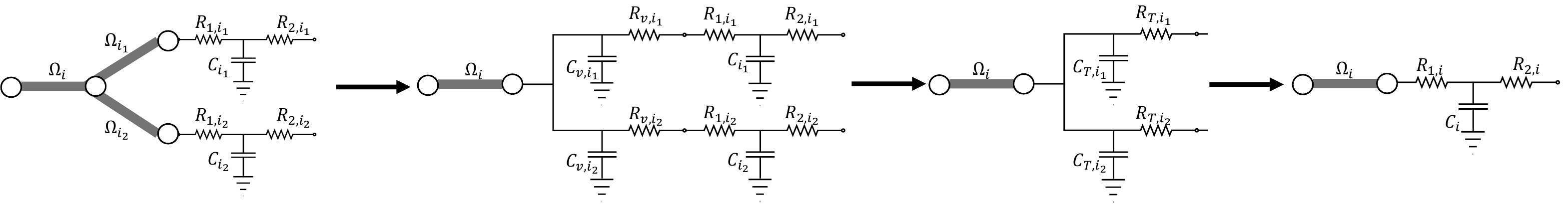}
    \caption{Schematic illustration for lumping the Windkessel model parameters~\cite{epstein2015reducing}.}
    \label{fig:rcr_lumping}
\end{figure}

Since each junction connects one parent vessel segment and two child segments, following \cite{epstein2015reducing}, the outlet Windkessel models associated with the child vessels can be systematically aggregated into an equivalent Windkessel model for the parent segment. For a segment $\Omega_i$ with $v_i \in \mathcal{V}^{\mathrm{conj}}$, we consider its child vessels $\Omega_{i_1}$ and $\Omega_{i_2}$, each equipped with a three-element Windkessel (RCR) model, and first reduce them to an equivalent 0-D model (see Fig.~\ref{fig:rcr_lumping}). The resulting total resistance $R_{T,i_k}$ and compliance $C_{T,i_k}$ are given by
\begin{equation}
\begin{aligned}
R_{T,i_k} &= R_{1,i_k} + R_{2,i_k} + R_{v,i_k}, \\
C_{T,i_k} &= \frac{C_{v,i_k} R_{1,i_k} + C_{v,i_k} R_{2,i_k} + C_{i_k} R_{2,i_k} + C_{v,i_k} R_{v,i_k}}{R_{1,i_k} + R_{2,i_k} + R_{v,i_k}}, \qquad k=1,2,
\end{aligned}
\end{equation}
where $R_{v,i_k}$ and $C_{v,i_k}$ denote the effective resistance and compliance of the vessel segment, defined as
\begin{equation}
\begin{aligned}
R_{v,i_k} &= 2 (\xi + 2)\pi \mu \int_0^{L_{i_k}} \frac{1}{A_{i_k}^2} \, dx, \\
C_{v,i_k} &= \frac{1}{\rho_i} \int_0^{L_{i_k}} \frac{A_{i_k}}{c_{i_k}^2} \, dx, \qquad k=1,2,
\end{aligned}
\end{equation}
where $\xi=9$ is the polynomial order of the velocity profile, $\mu = 4\;\text{mPa}\cdot\text{s}$ the blood dynamic viscosity, following the assumptions in~\cite{epstein2015reducing}.

The two resulting 0-D models are then combined into a single equivalent three-element Windkessel model for the parent segment $\Omega_i$, characterized by parameters $R_{1,i}$, $R_{2,i}$, and $C_i$. The proximal resistance $R_{1,i}$ is chosen as the characteristic impedance $Z_i$ to minimize wave reflections. The equivalent parameters are given by
\begin{align}
R_{1,i} &= Z_i, \\
R_{2,i} &= \frac{R_{T,i_1} R_{T,i_2}}{R_{T,i_1} + R_{T,i_2}} - R_{1,i}, \\
C_i &= C_{T,i_1} + C_{T,i_2},
\end{align}
where the characteristic impedance is defined as
\begin{equation}
Z_i = \frac{\rho_i c_i}{A_{d,i}},
\end{equation}
and $c_i$ denotes the pulse wave velocity (PWV), given by
\begin{equation}
c_i = \sqrt{\frac{\beta_i}{2 \rho_i A_{d,i}}} \, A_{d,i}^{1/4}.
\end{equation}

\section{Bifurcation-Based Domain Decomposition}

\subsection{Rationale for Using Bifurcation Units}

The original formulation of the 1D vascular system is highly coupled, as the entire network must be solved simultaneously to satisfy the junction conditions in~\eqref{eq:conjunction} and the inlet and outlet boundary conditions in~\eqref{eq:inlet} and~\eqref{eq:outlet}. Although numerical solvers are available for this purpose, they can be computationally expensive and challenging to implement, particularly for large-scale vascular networks.

Motivated by recent advances in operator learning, we aim to develop a model that can efficiently approximate the solution in a data-driven manner. A straightforward approach would be to treat the entire vascular network as a single training unit. However, this is impractical, as generating training data for complete vessel systems is costly and requires extensive variations in geometry and physiological parameters (e.g., across different individuals with varying age, gender, height, and weight). 

Instead, we exploit the observation that most vascular systems can be represented by tree-like structures and decompose them into smaller units that share common topological patterns. If an operator network can be trained to map unit-specific parameters to local solution fields for such representative units, then the global solution can be reconstructed by assembling the solutions over all units.

A natural decomposition strategy would be to consider individual vessel segments as basic units. However, for segments other than the root segment, the inlet boundary conditions are not directly available. In particular, there is no explicit flow-splitting relation that determines how the flow from a parent vessel is distributed among its child segments, even though there are conjunction conditions describing the equality between the flux out from the parent and the flux summed over the child segments, see~\eqref{eq:conjunction}.

To address this issue, we instead adopt bifurcation structures as the fundamental units for domain decomposition, which we refer to as bifurcation units (BUs) hereafter. With this choice, the inlet boundary condition for a child unit can be directly obtained from its parent, since the corresponding interface is contained within the parent domain. For outlet boundary conditions, we employ Windkessel models to represent downstream effects. Using the parameter lumping strategy described in Section~\ref{sec:rcr_lumping}, the distal vascular beds can be recursively aggregated into equivalent Windkessel models, thereby providing consistent and tractable boundary conditions for each BU.

We now formally introduce the domain decomposition. Let the complete vascular domain be defined as $\Omega := \bigcup_{i=1}^{N} \Omega_i.$ Then, the domain can be decomposed into a collection of BUs as
\begin{equation}
\Omega = \bigcup\limits_{v_i \in \mathcal{V}^{\mathrm{conj}}} \mathcal{B}_i,
\end{equation}
where each BU $\mathcal{B}_i$ is defined by $\mathcal{B}_i = \Omega_i \cup \Omega_{i_1} \cup \Omega_{i_2}$, with $i_1$ and $i_2$ denoting the indices of the two child segments of the parent segment $\Omega_i$. Thus, the points in $\mathcal{V}^\text{conj}$ could naturally be a one-to-one correspondence that could also be the representatives of the BUs.

\subsection{Multi-Level Schwarz Waveform Relaxation by BU with Parameter Lumping}

\begin{figure}
    \centering
    \includegraphics[width=\linewidth]{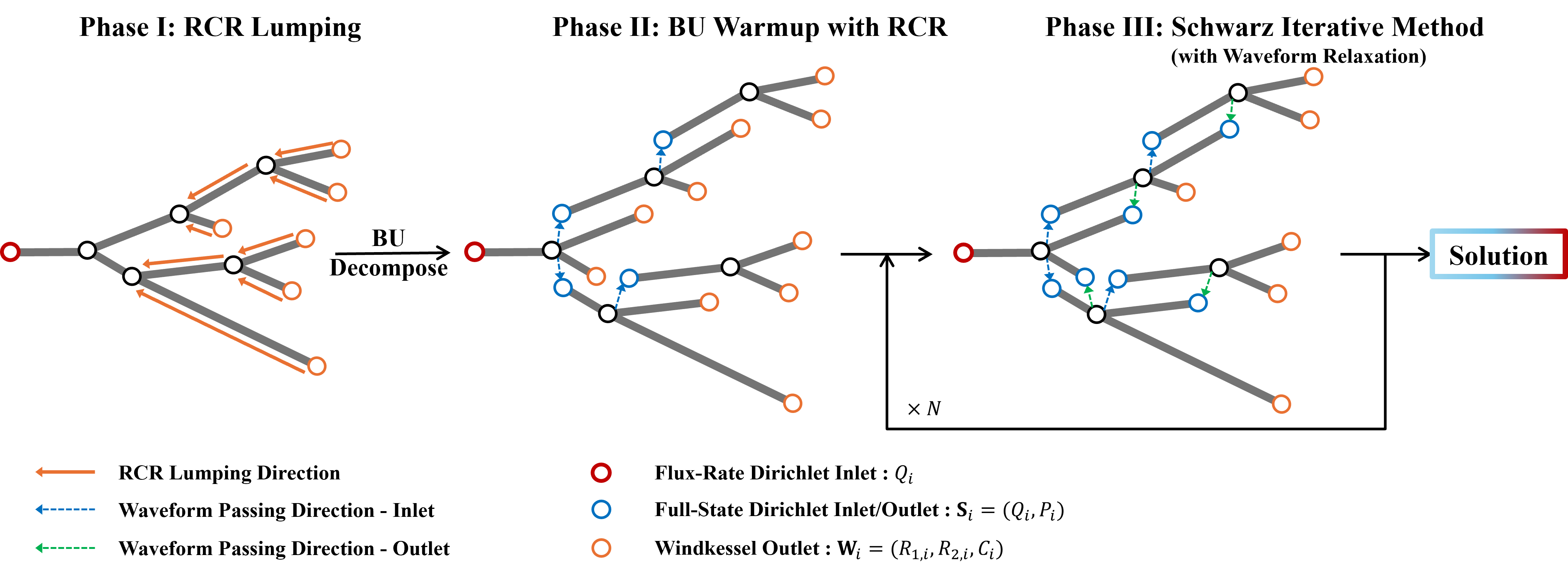}
    \caption{Illustration of the three phases of Algorithm~\ref{alg:main}: (I) Bottom-up parameter aggregation is performed on the original vascular tree, where terminal nodes (orange) are equipped with Windkessel boundary conditions and equivalent lumped Windkessel models are constructed for intermediate nodes (black) following Section~\ref{sec:rcr_lumping}. (II) The vessel tree is decomposed into BUs, and a top-down initialization is performed. The root BU is solved with the prescribed-flow Dirichlet inlet condition $Q$, while downstream BUs receive full-state Dirichlet inlet conditions $\mathcal{S}=(Q,P)$ from their parent BUs (blue), with dashed blue arrows indicating the information transfer direction. The Windkessel model obtained from Phase I is used for the outlet. (III) Schwarz waveform relaxation iteratively updates the full-state Dirichlet interface conditions between neighboring BUs (green and blue arrows), while terminal BUs retain their Windkessel outlet models.
}
    \label{fig:algorithm}
\end{figure}

\begin{algorithm}[ht]
\caption{The main algorithm. }
\begin{algorithmic}[1]
\State \textbf{Input:} Network topology, Inflow profile $Q_{\text{in}}$, Windkessel Parameters of leaf nodes.
\Statex \Comment{\textbf{Phase I: Parameter Aggregation (Bottom-Up)}}
\For{$j=J$ \textbf{to} $2$}
    \For{$i^{th}$ indexing the nodes in depth $j$}
        \State $\mathbf{W}_i \gets$ Compute $(R_{1,i}, R_{2,i}, C_i)$ via parameter lumping (Section~\ref{sec:rcr_lumping})
    \EndFor
\EndFor

\Statex \Comment{\textbf{Phase II: Warm-Start Initialization (Top-Down)}}
\State $\mathbf{S}_{\text{in},1}^{(0)}$ $\leftarrow$ $(Q_{\text{in}}, /)$
\For{$j=1$ \textbf{to} $J$}
    \For{$i$ indexing the nodes in depth $\mathcal{V}^j \cap \mathcal{V}^\text{conj}$}
        \State Solve $\mathcal{B}_i$ with $(\mathbf{S}_{\mathrm{in},i}^{(0)}, \mathbf{W}_{i_1},\mathbf{W}_{i_2})$ to obtain $\mathbf{S}_{i,i}^{(0)}, \mathbf{S}_{i,i_1}^{(0)}, \mathbf{S}_{i,i_2}^{(0)}$
        \State Set $\mathbf{S}_{\mathrm{in},i_1}^{(0)} \gets \mathbf{S}_{i,i_1}^{(0)}(0,t)$, \quad $\mathbf{S}_{\mathrm{in},i_2}^{(0)} \gets \mathbf{S}_{i,i_2}^{(0)}(0,t)$
    \EndFor
\EndFor

\Statex \Comment{\textbf{Phase III: Multiplicative Schwarz Waveform Relaxation}}
\State $k \gets 1$, 
\While{$\|\mathbf{S}^k - \mathbf{S}^{k-1}\| > \epsilon$}
    \State $\mathbf{S}_{\text{in},1}^{(k)}$ $\leftarrow$ $(Q_{\text{in},1},/)$
    \For{$j=1$ \textbf{to} $J - 1$}
        \For{$i$ indexing the nodes in depth $\mathcal{V}^j \cap \mathcal{V}^\text{conj}$}
            \State $\mathbf{S}_{\mathrm{out},i_1}^{(k)}\gets\mathbf{S}_{i_1,i_1}^{(k-1)}(L_{i_1},t)$, $\mathbf{S}_{\mathrm{out},i_2}^{(k)}\gets\mathbf{S}_{i_2,i_2}^{(k-1)}(L_{i_2},t)$
            \State Solve $\mathcal{B}_i$ with $\big(\mathbf{S}_{\mathrm{in},i}^{(k)}, \mathbf{S}_{\mathrm{out},i_1}^{(k)}, \mathbf{S}_{\mathrm{out},i_2}^{(k)}\big)$ to obtain $\mathbf{S}_{i,i}^{(k)}, \mathbf{S}_{i,i_1}^{(k)}, \mathbf{S}_{i,i_2}^{(k)}$
            \State Set $\mathbf{S}_{\mathrm{in},i_1}^{(k)} \gets \mathbf{S}_{i,i_1}^{(k)}(0,t)$, $\mathbf{S}_{\mathrm{in},i_2}^{(k)} \gets \mathbf{S}_{i,i_2}^{(k)}(0,t)$
            \State $\mathbf{S}_i^{(k)} \gets \mathbf{S}_{i,i}^{(k)}$
        \EndFor
    \EndFor
    
    \For{$i$ indexing the nodes in $\mathcal{V}^J \cap \mathcal{V}^\text{conj}$}
        \State Solve $\mathcal{B}_i$ with $(\mathbf{S}_{\mathrm{in},i}^{(k)}, \mathbf{W}_{i_1},\mathbf{W}_{i_2})$ to obtain $\mathbf{S}_{i,i}^{(k)}, \mathbf{S}_{i,i_1}^{(k)}, \mathbf{S}_{i,i_2}^{(k)}$
        \State $\mathbf{S}_i^{(k)} \gets \mathbf{S}_{i,i}^{(k)}$, $\mathbf{S}_{i_1}^{(k)} \gets \mathbf{S}_{i,i_1}^{(k)}$, $\mathbf{S}_{i_2}^{(k)} \gets \mathbf{S}_{i,i_2}^{(k)}$
    \EndFor
\State $k \gets k + 1$
\EndWhile
\State \textbf{Output:} $\{\mathbf{S}_i^{(k)}\}_{i=1}^N$\Comment{Solution in all domains}
\end{algorithmic}
\label{alg:main}
\end{algorithm}

Based on the BU decomposition, we formulate the iterative coupling across units with Schwarz waveform relaxation (SWR). Unlike classical Schwarz methods that exchange pointwise-in-time boundary values at each time step, SWR exchanges entire time-dependent waveforms at subdomain interfaces, each BU is solved over the full time window $[0, T_{\mathrm{cardiac}}]$, where $T_{\mathrm{cardiac}}$ denotes the cardiac cycle period, and the resulting interface waveforms are passed to neighboring units as updated boundary conditions for the next iteration. The multiplicative variant updates BUs sequentially within each iteration so that downstream units immediately benefit from the most recent upstream solution.

The proposed BU-based algorithm consists of three phases. Since 6 types of boundary-condition configurations are involved throughout the procedure, we summarize their usage in each phase in Table~\ref{tab:bu_types} for clarity. Fig.~\ref{fig:algorithm} illustrates the overall workflow, including parameter aggregation, warm-start initialization, and multiplicative Schwarz waveform relaxation. We first introduce a level structure on the binary tree, where level $j$ consists of all nodes whose distance from the root is $j$. We denote the maximum depth of the tree by $J+1$, where the root node corresponds to level $j=0$. Accordingly, the node set can be partitioned into disjoint subsets as
\begin{equation}
    \mathcal{V} = \mathcal{V}^{0} \cup \mathcal{V}^{1} \cup \cdots \cup \mathcal{V}^{J+1},
\end{equation}
where $\mathcal{V}^{j}$ denotes the collection of nodes at level $j$, for $j=0,\ldots,J+1$. For convenience, we denote the local solution variables for each vessel segment $\Omega_i$ by 
\begin{equation}
    \mathbf{S}_i = (Q_i, P_i).
\end{equation}

In the first phase, a bottom-up parameter aggregation procedure is performed to construct effective outlet boundary conditions for all BUs $\mathcal{B}_i$, where $v_i \in \mathcal{V}^{\mathrm{conj}}$. Specifically, the aggregation is carried out for nodes from level $J$ down to level $2$. Nodes at level $J+1$ are terminal/leaf nodes indicated as Windkessel outlet (orange in Fig.~\ref{fig:algorithm}) and already possess prescribed outlet boundary conditions, while nodes at level $1$ do not require aggregation since their associated BU connects directly to level $2$. Therefore, for each intermediate node $v_i$ (black in Fig.~\ref{fig:algorithm}) at levels $j=J,\ldots,2$, corresponding to segment $\Omega_i$, we aggregate the Windkessel parameters of its child segments following the procedure described in Section~\ref{sec:rcr_lumping}. The effect of downstream vessels is then incorporated through the aggregated Windkessel parameters, allowing each local decomposed BU to have its own outlet model. 

In the second phase, a top-down solution procedure is performed over the decomposed BUs. The procedure starts from the root BU $\mathcal{B}_1$, whose inlet boundary condition at the root node is prescribed as a flux-Dirichlet inlet condition $Q$ (red node in Fig.~\ref{fig:algorithm}). Together with the effective Windkessel outlet boundary condition (orange node in Fig.~\ref{fig:algorithm}) obtained from Phase I, the solution of the root BU can be computed using either the neural operator or the numerical solver. The solution is then propagated recursively from level $j=1$ to $j=J$. Specifically, for each conjunction node $v_i \in \mathcal{V}^{j}\cap\mathcal{V}^{\mathrm{conj}}$, the associated BU $\mathcal{B}_i=\Omega_i\cup\Omega_{i_1}\cup\Omega_{i_2}$ is solved using its inlet and outlet boundary conditions by a neural operator or numerical method. The resulting solutions $\mathbf{S}_{i,i}^{(0)}$, $\mathbf{S}_{i,i_1}^{(0)}$, and $\mathbf{S}_{i,i_2}^{(0)}$ (where the first index denotes the BU and the second index denotes the corresponding vessel segment) are defined on $\Omega_i$, $\Omega_{i_1}$, and $\Omega_{i_2}$, respectively. For the child BUs $\mathcal{B}_{i_1}$ and $\mathcal{B}_{i_2}$ at level $j+1$, the inlet boundary conditions at the full-state Dirichlet interfaces (blue nodes in Fig.~\ref{fig:algorithm}) are inherited from the corresponding parent BU solutions at the shared conjunction nodes. The direction of this information transfer is indicated by the dashed blue arrows in Fig.~\ref{fig:algorithm}, and the inlet states are assigned as
\begin{equation}
    \mathbf{S}_{\mathrm{in},i_1}^{(0)}
    \gets
    \mathbf{S}_{i,i_1}^{(0)}(0,t),
    \qquad
    \mathbf{S}_{\mathrm{in},i_2}^{(0)}
    \gets
    \mathbf{S}_{i,i_2}^{(0)}(0,t).
    \label{eq:conj_inlet}
\end{equation}

In the third phase, a multiplicative Schwarz waveform relaxation is applied to iteratively refine the initial solution obtained in Phase II. At each iteration $k$, the BUs are updated sequentially in a top-down manner using the most recent interface information exchanged between neighboring units. For the root BU, the inlet boundary condition remains the prescribed flow-rate Dirichlet condition $Q$. The full-state Dirichlet outlet conditions (blue in Fig.~\ref{fig:algorithm}) are updated using the solutions of its child BUs from the previous iteration. The direction of this information transfer is indicated by the dashed green arrows in Fig.~\ref{fig:algorithm}, and the outlet states are assigned as
\begin{equation}
    \mathbf{S}_{\mathrm{out},i_1}^{(k)}
    \gets
    \mathbf{S}_{i_1,i_1}^{(k-1)}(L_{i_1},t),
    \qquad
    \mathbf{S}_{\mathrm{out},i_2}^{(k)}
    \gets
    \mathbf{S}_{i_2,i_2}^{(k-1)}(L_{i_2},t).
    \label{eq:conj_outlet}
\end{equation}
With these updated boundary conditions, the solution is propagated recursively from level $j=1$ to $j=J$ following the same top-down procedure as in Phase II. For each BU $\mathcal{B}_i$, the local solutions $\mathbf{S}_{i,i}^{(k)}$, $\mathbf{S}_{i,i_1}^{(k)}$, and $\mathbf{S}_{i,i_2}^{(k)}$ are obtained on $\Omega_i$, $\Omega_{i_1}$, and $\Omega_{i_2}$, respectively. During this propagation, as illustrated in Fig.~\ref{fig:algorithm}, the full-state Dirichlet inlet conditions of the child BUs $\mathcal{B}_{i_1}$ and $\mathcal{B}_{i_2}$ are updated from the parent BU solutions according to \eqref{eq:conj_inlet}, with the corresponding information transfer indicated by the dashed blue arrows. Meanwhile, the full-state Dirichlet outlet conditions are corrected using the children BU solutions from the previous iteration according to \eqref{eq:conj_outlet}, as indicated by the dashed green arrows. This iterative exchange of interface waveforms continues until the solutions of neighboring BUs become consistent. For terminal BUs, the outlet boundary condition is retained as the prescribed Windkessel outlet model.

This Phase III iterative procedure enforces consistency across overlapping subdomains and progressively reduces interface mismatches. The iteration terminates once a prescribed convergence criterion is satisfied, yielding a globally consistent solution over the entire vascular network. As shown empirically in Section~\ref{sec:experiments}, the residual decreases monotonically, so in practice we run the iteration for a fixed budget of $k$ iterations, which is chosen such that the difference between two successive iterates is sufficiently small, indicating convergence of the interface waveforms and the overall solution. Additional convergence results for networks of intermediate size are provided in \ref{app:swr_convergence_size}.

A key advantage of this framework is that the local solves within each BU can be efficiently performed using a trained operator network, which replaces expensive high-fidelity numerical simulations. As a result, the overall method achieves a balance between computational efficiency and physical fidelity, making it well-suited for large-scale and subject-specific vascular modeling. In the next section, we will describe the details of the operator network.

\begin{remark}\label{rmk:bc}
    For the BU containing the origin node, the inlet boundary condition is prescribed only in terms of the flow rate $Q$. In contrast, for all other BUs, the inlet boundary condition is imposed on the full state variable $\mathbf{S} = (Q, P)$, including both flow rate and pressure (equivalently, the cross-sectional area $A$ via the tube law in~\eqref{eq:wall}).
    It is also worth noting that imposing both $Q$ and $P$ is due to the need to transfer the two Riemann invariants that carry information between parent and child segments in 1D Navier–Stokes models for vascular networks~\cite{formaggia2003one}. 
    These quantities are not fixed a priori; instead, they are iteratively updated during computation through the coupling correction steps across segments, ensuring consistency of the final solution pair $(Q,P)$ with the governing system.
    Moreover, from our experimental experience, transferring only a single variable in the alternating correction step generally led to numerical instability and eventual divergence of the full vascular network simulation.
\end{remark}
\section{Operator Network for Bifurcation Unit}

Algorithm~\ref{alg:main} requires solving each bifurcation unit (BU) repeatedly. Each BU is solved once during the top-down warm-start in Phase~II, and then at every SWR iteration in Phase~III. By replacing the numerical BU solver with the learned BUFormer operator (see Section~\ref{sec:architecture}), each local solver reduces to a single forward pass, and the overall algorithm can reach a converged solution faster than a conventional 1D numerical simulation of the whole network. Given a BU and its physical parameters and boundary conditions, the operator predicts the local solution fields directly. A key advantage of this design is that the operator entirely focuses on mapping local unit parameters to the solution, thus has no dependence on the global network topology. The same trained operator therefore can be applied, without any retraining or architectural modification, to any vascular tree that can be decomposed into BUs, including networks with a different number of segments or with missing vessel segments. Training is performed once offline and the operator is reused throughout the algorithm. This section describes the input-output formulation of the operator (Section~\ref{sec:bu_io}), the network architecture (Section~\ref{sec:architecture}), and the optimization objective (Section~\ref{sec:training}).

\subsection{BU Input-Output Map}
\label{sec:bu_io}

Consider a BU $\mathcal{B}_i = \Omega_i \cup \Omega_{i_1} \cup \Omega_{i_2}$. We seek a parametric operator $\mathcal{G}_\theta$ of the form
\begin{equation}
    \mathcal{G}_\theta \;:\; (\boldsymbol{s}_i,\; g_{\mathrm{in},i},\; g_{\mathrm{out},i_1},\; g_{\mathrm{out},i_2}) \;\longmapsto\; (\mathbf{S}_i,\; \mathbf{S}_{i_1},\; \mathbf{S}_{i_2}),
    \label{eq:bu_operator}
\end{equation}
where the inputs are:
\begin{itemize}
    \item $\boldsymbol{s}_i \in \mathbb{R}^{d_s}$: a vector of scalar geometric and physical parameters of the three vessel segments (e.g., vessel radius, length, wall stiffness $\beta$, reference pressure, cardiac cycle period);
    \item $g_{\mathrm{in},i}$: the inlet boundary condition, either the prescribed root inflow $Q_{\mathrm{in},1}(t)$ or the full state trace $\mathbf{S}_{i}(0,t)$ inherited from the parent unit via~\eqref{eq:conj_inlet};
    \item $g_{\mathrm{out},i_k}$ for $k=1,2$: the outlet boundary condition for each daughter segment, either the lumped resistances and compliance parameters $(R_{1,i_k},R_{2,i_k}, C_{T,i_k})$ or the Dirichlet outlet trace $\mathbf{S}_{\mathrm{out},i_k}(t)$ obtained from downstream units via~\eqref{eq:conj_outlet};
\end{itemize}
and the output $\mathbf{S}_j = \mathbf{S}_j(x_m, t)$ for $j \in \{i, i_1, i_2\}$ is the solution field evaluated at a set of points $\{x_m\}$ discretizing the segment $\Omega_j$ over one cardiac cycle. Since the inlet admits two types of boundary conditions and the pair of outlets admits three possible combinations, there are $2\times 3 = 6$ possible configurations for a single BU. Table~\ref{tab:bu_types} enumerates all six types and indicates which ones arise during Algorithm~\ref{alg:main}.

\begin{table}[ht!]
\centering
\footnotesize
\caption{The six possible boundary condition (BC) configurations for a bifurcation unit, determined by the inlet type (Dirichlet by flux-rate $Q$ for the root BU, or Dirichlet by full-state trace $\mathbf{S}$ for non-root BUs) and by the outlet types of the two daughter segments, which admit three possible combinations of Windkessel or Dirichlet conditions by $Q$ or $\mathbf{S}$. Types~1--3 occur only at the root unit; types~4--6 occur at all other units. Types~1,4 can occur in Phase II, while Types~2-6 can occur in Phase III.}
\begin{tabular}{c c c c}
\hline
\textbf{Type} & \textbf{Inlet} & \textbf{Outlets} & \textbf{Phase}\\
\hline
1 & Flux-Rate $Q$ & Windkessel + Windkessel & II \\
2 & Flux-Rate $Q$ & Windkessel + Full-State $\mathbf{S}$ & III \\
3 & Flux-Rate $Q$ & Full-State $\mathbf{S}$  + Full-State $\mathbf{S}$ & III \\
4 & Full-State $\mathbf{S}$ & Windkessel + Windkessel & II \& III\\
5 & Full-State $\mathbf{S}$ & Windkessel + Full-State $\mathbf{S}$ & III\\
6 & Full-State $\mathbf{S}$ & Full-State $\mathbf{S}$  + Full-State $\mathbf{S}$ & III\\
\hline
\end{tabular}
\label{tab:bu_types}
\end{table}

$\mathcal{G}_\theta$ only accepts local information of the BU and has no information about the position of the unit in the vascular tree or the topology of the surrounding network. In addition, traditional 1D solvers typically require several cyclic time windows for the blood flow to stabilize and reach a periodic solution. In contrast, our operator directly predicts the converged solution fields over the stabilized cardiac cycle with a single forward pass, thus allowing to achieve significant speedup at the BU level.

\subsection{Network Architecture}
\label{sec:architecture}

\begin{figure}[t!]
    \centering
    \includegraphics[width=\linewidth]{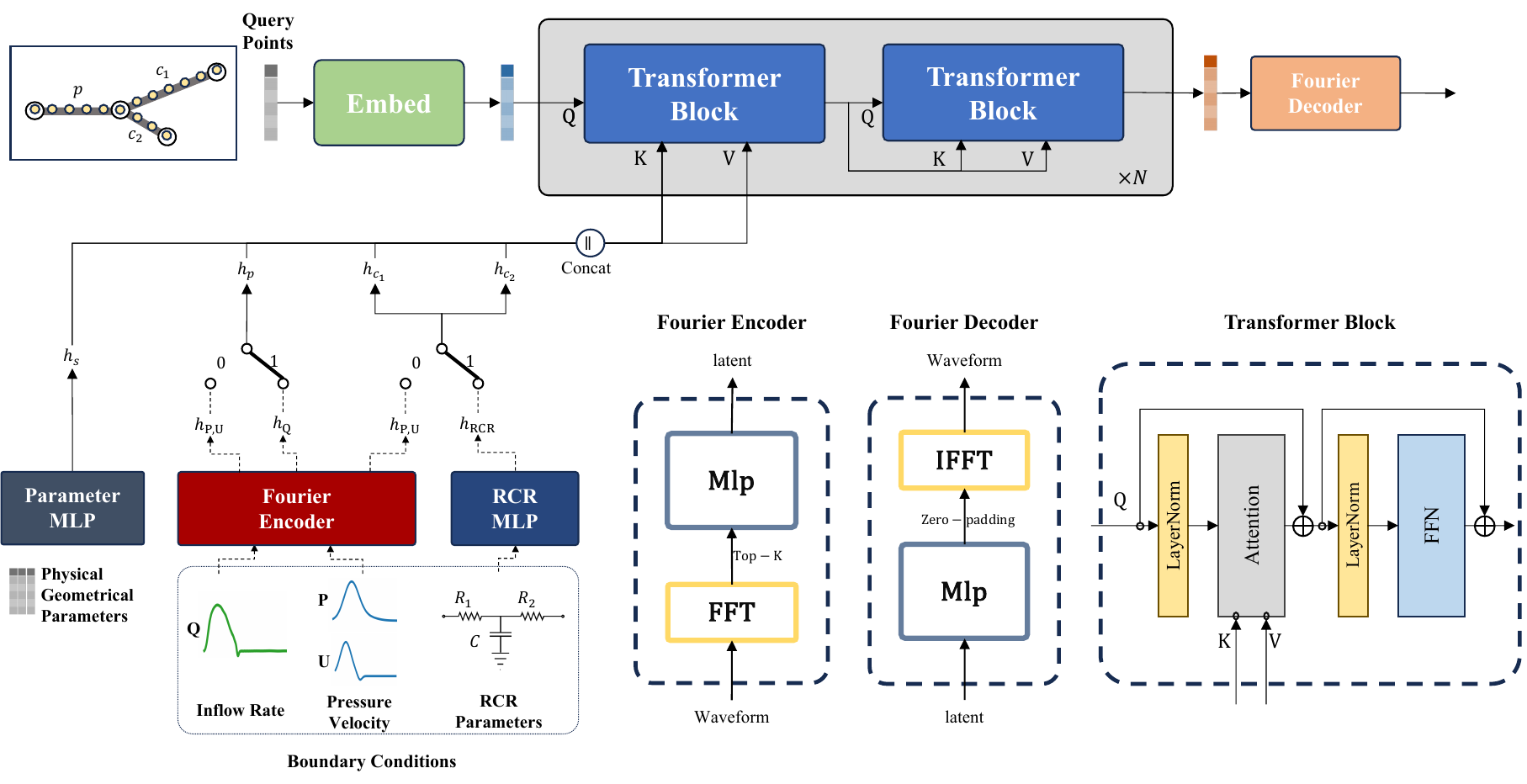}
    \caption{Overview of the BUFormer encoder-decoder architecture. The encoder produces four tokens from the BU inputs. The decoder queries cross-attend to these tokens through $N$ Transformer blocks and are converted to temporal waveforms via a Fourier Decoder.}
    \label{fig:architecture}
\end{figure}

Based on the transformer architecture, we propose BUFormer (Fig.~\ref{fig:architecture}) as the neural solver for BUs, following an encoder-decoder design. The encoder maps each group of BU inputs into tokens in $\mathbb{R}^h$. A Parameter MLP produces a token $h_s$ from the scalar geometric and physical parameters $\boldsymbol{s}_i$, a Fourier Encoder (described below) produces a token $h_p$ from the inlet boundary waveform $g_{\mathrm{in},i}$, and a mixture of Windkessel parameters MLP and Fourier Encoder produces tokens $h_{c_1}$ and $h_{c_2}$ for the two outlet boundary conditions $g_{\mathrm{out},i_1}$ and $g_{\mathrm{out},i_2}$, respectively. These tokens are assembled into a context set $\boldsymbol{H} = [h_s,\, h_p,\, h_{c_1},\, h_{c_2}] \in \mathbb{R}^{4 \times h}$, serving as the key-value pairs for cross-attention. Then, query points are defined along the spatial coordinates of the three vessel segments. Each query is processed through $N$ Transformer blocks that cross-attend to the context set $\boldsymbol{H}$, enabling the aggregation of information from all input groups, capturing the coupling between boundary conditions and vessel parameters. The resulting embedding is then decoded by a Fourier Decoder for the output solution fields.

Since the hemodynamic solution is quasi-periodic over the cardiac cycle, the spectral representations provide a compact and expressive basis. Thus, we exploit this structure in both the encoder and decoder. The Fourier Encoder maps an input waveform $f(t) \in \mathbb{R}^{N_t}$ to a latent token by
\begin{equation}
    \mathrm{FourierEnc}(f) = \mathrm{MLP}\!\left(\mathrm{LN}\!\left(\bigl[\operatorname{Re}(\hat{f}_0),\, \ldots,\, \operatorname{Re}(\hat{f}_{K-1}),\, \operatorname{Im}(\hat{f}_0),\, \ldots,\, \operatorname{Im}(\hat{f}_{K-1})\bigr]\right)\right) \in \mathbb{R}^h,
    \label{eq:fourier_enc}
\end{equation}
where $\hat{f} = \mathrm{FFT}(f)$, only the first $K$ low-frequency modes are retained, and $\mathrm{LN}$ denotes layer normalization. 

Symmetrically, the Fourier Decoder reconstruct a temporal waveform from each query's latent vector $\boldsymbol{z} \in \mathbb{R}^h$ via
\begin{equation}
    \mathrm{FourierDec}(\boldsymbol{z}) = \mathrm{IFFT}\!\left(\mathrm{ZeroPad}_{N_t}\!\bigl(\mathrm{MLP}(\boldsymbol{z})\bigr)\right) \in \mathbb{R}^{N_t},
    \label{eq:fourier_dec}
\end{equation}
where the MLP outputs $K$ complex coefficients that are zero-padded to the full spectral resolution before the inverse transform. This parameterization yields a compact and smooth temporal representation with far fewer degrees of freedom than a pointwise prediction.

Each Transformer decoder block performs cross-attention followed by a feed-forward network
\begin{equation}
\begin{aligned}
    \boldsymbol{z}' &= \boldsymbol{z} + \mathrm{Attn}(\mathrm{LN}(\boldsymbol{z}),\; \boldsymbol{H},\; \boldsymbol{H}), \\
    \boldsymbol{z}'' &= \boldsymbol{z}' + \mathrm{FFN}(\mathrm{LN}(\boldsymbol{z}')),
\end{aligned}
\label{eq:transformer_block}
\end{equation}
where $\boldsymbol{H}$ is the encoder token matrix defined above, and $\mathrm{Attn}(Q, K, V)$ is multi-head attention. This is repeated for $N$ blocks.

As described in Section~\ref{sec:bu_io}, the inlet and each outlet of a BU can carry different boundary condition types depending on the unit's position in the vascular tree. The encoder handles this via a mask-based switch, where for each boundary, a binary mask $m \in \{0, 1\}$ selects between two encoding branches as $h = m \cdot h_{\mathrm{A}} + (1-m) \cdot h_{\mathrm{B}}$. At the inlet, the mask routes either a flow-rate waveform (root BU) or a full state trace (non-root BU) to the appropriate Fourier Encoder. At each outlet, the mask selects between the RCR MLP for Windkessel parameters and the Fourier Encoder for Dirichlet traces. This unified design handles all boundary condition combinations in Table~\ref{tab:bu_types} without any architectural change, and enables efficient batching over different boundary condition types. Specific hyperparameter values for the two experimental settings are reported in Appendix~\ref{app:hyperparams}.

\subsection{Optimization Objective}
\label{sec:training}

The network parameters $\theta$ are optimized by minimizing the relative $L^2$ error between the predicted and reference solution fields. For a batch of $B$ samples, the loss is
\begin{equation}
    \mathcal{L} = \frac{1}{B}\sum_{b=1}^{B} \frac{\|\hat{\mathbf{S}}^{(b)} - \mathbf{S}^{(b)}\|_2}{\|\mathbf{S}^{(b)}\|_2},
    \label{eq:training_loss}
\end{equation}
where $\hat{\mathbf{S}}^{(b)}$ and $\mathbf{S}^{(b)}$ denote the predicted and reference solutions for the $b$-th sample, concatenated over all three segments, spatial history points, fields, and time steps. The same relative $L^2$ metric is used to evaluate accuracy in all experiments reported in Section~\ref{sec:experiments}.

\section{Experimental Results}
\label{sec:experiments}

\subsection{Experimental Setup}

We evaluate the proposed framework at two levels. First, we assess the accuracy of the BU operator $\mathcal{G}_\theta$ as a local surrogate on individual bifurcation units (BUs). Second, we evaluate its effectiveness as a component inside the full SWR-based network solver (Algorithm~\ref{alg:main}).

\paragraph{Datasets}

\begin{figure*}[t!]
    \centering
    \includegraphics[width=\textwidth]{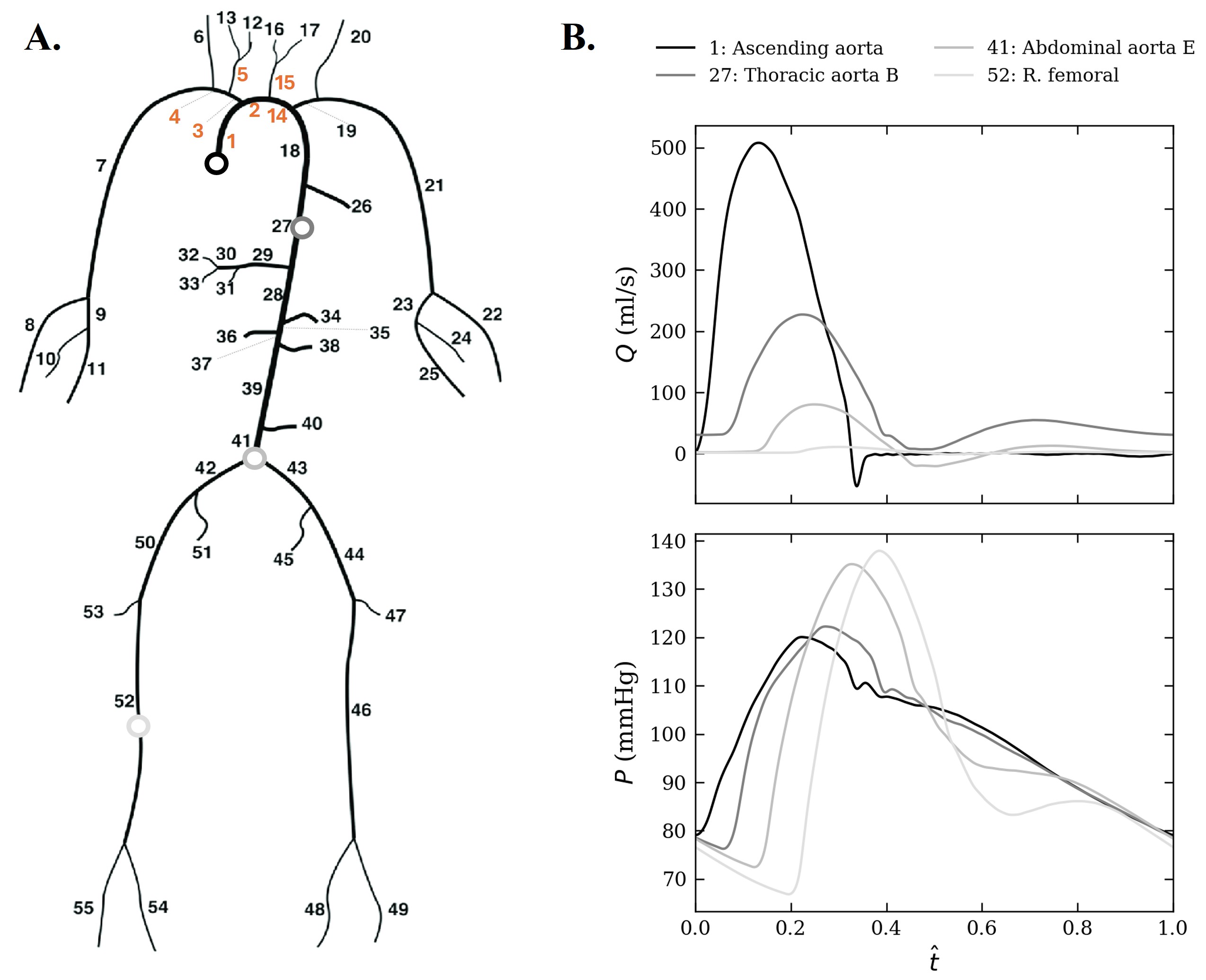}
    \caption{Overview of the 55-segment arterial tree model~\cite{Willemet2015}. (A)~Schematic of the full arterial network with segment numbering. The 7-segment subset used for initial validation (segments 1--5, 14, 15) is highlighted in orange. Gray circles indicate the measurement locations corresponding to the waveforms in~(B). (B)~Exemplar flow rate $Q$ (top) and pressure $P$ (bottom) waveforms over one cardiac cycle at four locations along the aortic-to-femoral pathway, illustrating the characteristic attenuation of flow and steepening of the pressure pulse with increasing distance from the aortic root. The normalized time is $\hat{t} = t / T$.}
    \label{fig:55seg_demo}
\end{figure*}

We adopt the 55-segment arterial tree model of \cite{Willemet2015} as the primary vascular network (Fig.~\ref{fig:55seg_demo}). The baseline physical, material, geometric, and Windkessel parameters are listed in Table~\ref{tab:55segments} (Appendix~\ref{app:baseline_params}). As a smaller test case for demonstration, we also consider a 7-segment subset consisting of segments 1--5, 14, and 15 (marked with $\dagger$ in Table~\ref{tab:55segments} and highlighted in orange in Fig.~\ref{fig:55seg_demo}A) of the same tree. 

Following~\cite{Willemet2015}, we control arterial stiffness and lumen size through age-dependent multiplicative scaling factors (Table~\ref{tab:datasets-age_trajectory}). The wall stiffness parameter $\beta_i$ is related to the diastolic pulse wave velocity $c_i$ by evaluating the PWV expression at $A_i = A_{d,i}$, giving $\beta_i = 2\rho_i\, c_i^2 \sqrt{A_{d,i}}$. Following the empirical relation $c_i = a_i / D_{\mathrm{d},i}^{b}$ with $b = 0.3$~\cite{Reymond2009}, we obtain
\begin{equation}
    \beta_i = \frac{2\rho_i\, a_i^2 \sqrt{A_{d,i}}}{D_{d,i}^{2b}}.
    \label{eq:beta_from_a}
\end{equation}
Let $\eta_i$ and $\delta_i$ denote the stiffness and diameter scaling factors for segment~$i$, respectively, with values taken from Table~\ref{tab:datasets-age_trajectory} according to whether the vessel is an elastic or muscular artery. The scaled parameters are
\begin{equation}
    a_i = \eta_i \, a_i^\text{ref}, \qquad D_{d,i} = \delta_i \, D_{d,i}^\text{ref},
    \label{eq:scaling}
\end{equation}
where $a_i^\text{ref}$ and $D_{d,i}^\text{ref}$ are the baseline values from Table~\ref{tab:55segments}. The patient-specific stiffness $\beta_i$ is then obtained by substituting \eqref{eq:scaling} into~\eqref{eq:beta_from_a}.

For each model, a population of virtual subjects is created using continuously sampled physiological parameters, see Table~\ref{tab:datasets-age_trajectory} and~\ref{tab:datasets-indep_params}. For each virtual subject, a numerical simulation using the Nektar1D solver~\cite{Alastruey2012} provides the ground truth, and the subsequent decomposition-based solve (Phases~II and~III) in our proposed Algorithm~\ref{alg:main} yields one training sample per independent BU solver, including all intermediate solutions from the Schwarz iterations used for dataset generation ($k=20$ for the 7-segment model and $k=30$ for the 55-segment model), balancing solution quality with computational budget. This ensures the operator is trained on a sufficiently representative range of boundary condition inputs that it will encounter during inference. The 55-segment dataset contains 1,783 valid cases (approximately 1.49 million BU samples); the 7-segment dataset contains 1,819 valid cases (approximately 115,000 BU samples). Both are split 80/10/10 by case into training, validation, and test sets. The training pipeline is a one-time cost and is performed entirely offline using an HPC cluster. In our implementation, for the 55-segment test, the generation of the training dataset required approximately 35 hours on an HPC cluster with 200 CPU cores, followed by approximately 20 hours of neural operator training.

\begin{table}
\footnotesize
\centering
\caption{Parameters for constructing training dataset. Physiological scaling factors used to generate virtual subjects datasets. A virtual age $s \sim \mathcal{U}[25,75]$ is sampled uniformly and the stiffness and diameter scaling factors are obtained by piecewise-linear interpolation along the aging trajectory nodes in (a), following~\cite{Willemet2015}. Here $\eta_{i}$ and $\delta_{i}$ denote multiplicative scaling factors for the stiffness and diameter of segment $\Omega_i$, respectively, depending on whether the vessel is an elastic or muscular artery. Elastic arteries comprise the main aortic segments from the ascending aorta to the abdominal aorta, while muscular arteries include all remaining arterial segments~\cite{Willemet2015}. For ages below 50, diameter factors are decoupled from the trajectory and sampled uniformly in $[0.9, 1.0]$.}
\begin{tabular}{c c c c c}
\hline
\textbf{Age} & $\eta_{i}$ El. & $\eta_{i}$ Musc. & $\delta_{i}$ El. & $\delta_{i}$ Musc. \\
\hline
25 & 0.80 & 0.80 & [0.90,1.00] & [0.90,1.00] \\
30 & 1.00 & 1.00 & [0.90,1.00] & [0.90,1.00] \\
40 & 1.30 & 1.00 & [0.90,1.00] & [0.90,1.00] \\
50 & 1.60 & 1.15 & 1.20 & 1.21 \\
60 & 1.90 & 1.15 & 1.20 & 1.21 \\
70 & 2.25 & 1.30 & 1.40 & 1.21 \\
75 & 2.25 & 1.30 & 1.40 & 1.21 \\
\hline
\end{tabular}
\label{tab:datasets-age_trajectory}
\end{table}

\begin{table}
\footnotesize
\centering
\caption{Parameters for constructing training dataset. The parameters in this table are sampled independently. Cardiac period specifies the duration of a single cardiac cycle in seconds, bpm denotes beats per minute. Stroke volume specifies the flow volume the heart pumps into the system. Peripheral resistance provides the total resistance $R_T$ at each outlet node. Reference pressure models scenarios from normotensive to hypertensive systems, with a baseline normotensive diastolic pressure $P_{d}=12.67 \text{ kPa (95 mmHg)}$. Other than virtual age and cardiac period that are directly sampled, these scaling factors are multiplied to the baseline values specified in Tab. \ref{tab:55segments}.}
\begin{tabular}{l c l}
\hline
\textbf{Parameter} & \textbf{Range} & \textbf{Effect} \\
\hline
Virtual age (yr) & [25,75] & Age \\
Cardiac period (s) & $[0.75,\, 1.25]$ & Heart rate \\
 & & (48--80 bpm) \\
Stroke volume & $[0.80,\, 1.20]$ & Inflow amplitude \\
Periph.\ resistance & $[0.90,\, 1.10]$ & Outlet resistance \\
Ref.\ pressure & $[1.00,\, 1.40]$ & Diastolic pressure \\
\hline
\end{tabular}
\label{tab:datasets-indep_params}
\end{table}

Not all BC types from Table~\ref{tab:bu_types} are equally represented as we only run Phase II once but run Phase III iteratively. This imbalance is the result of our proposed algorithm. The root BU only produces type 1 at Phase II and type 3 at each Phase III iteration, while all other BUs contribute to types~(4,5,6). To prevent the operator from underfitting rare but structurally important configurations such as type 1 (which initiates the warm-start), we apply balanced sampling during training, reweighting each BC type to achieve uniform representation within each training batch.

\paragraph{Training Details}
The operator inputs and targets span several orders of magnitude and carry different units, so every field is normalized before training (details are addressed in Appendix~\ref{app:normalization}). The transformer is trained using the AdamW optimizer with a cosine annealing learning-rate schedule. The learning rate is initialized at $3\times10^{-4}$ and decayed to a minimum value of $1\times10^{-6}$, with a weight decay of $1\times10^{-4}$ and gradient clipping norm of 1.0. Boundary-condition types are balanced during training, and mixed-precision (FP16) training is adopted. The models are trained for 200 epochs and 80 epochs with batch sizes of 1024 and 800, respectively. More details can be found in Appendix~\ref{app:hyperparams}.

\paragraph{Performance Metrics}
We report the relative $L^2$ error defined in~\eqref{eq:training_loss}, computed separately for pressure and velocity, while the velocity is computed by $U = Q/A$. For readability, we also report agreement with the Nektar1D reference, defined as $100\,(1 - \varepsilon)$\,\%, where $\varepsilon$ is the mean relative $L^2$ error. For the full-network experiments, we additionally report speedups relative to the Nektar1D solver and the SWR convergence residual $\|\mathbf{S}^{(k)} - \mathbf{S}^{(k-1)}\|_2$, which measures how much the solution on the overlapping interface segments changes between successive Schwarz iterations.

\paragraph{Baselines}
We compare three solver configurations, namely (i)~\emph{Nektar1D}, where the full vascular network is solved by the 1D numerical solver, providing the ground-truth reference; (ii)~\emph{SWR Numerical}, which applies Algorithm~\ref{alg:main} with a numerical solver for each BU solve and is used for dataset generation; and (iii)~\emph{SWR Operator}, which applies Algorithm~\ref{alg:main} with the trained BU operator $\mathcal{G}_\theta$ replacing all local solves.

\subsection{Warm-Start Ablation}

\begin{figure}[ht!]
    \centering
    \includegraphics[width=\linewidth]{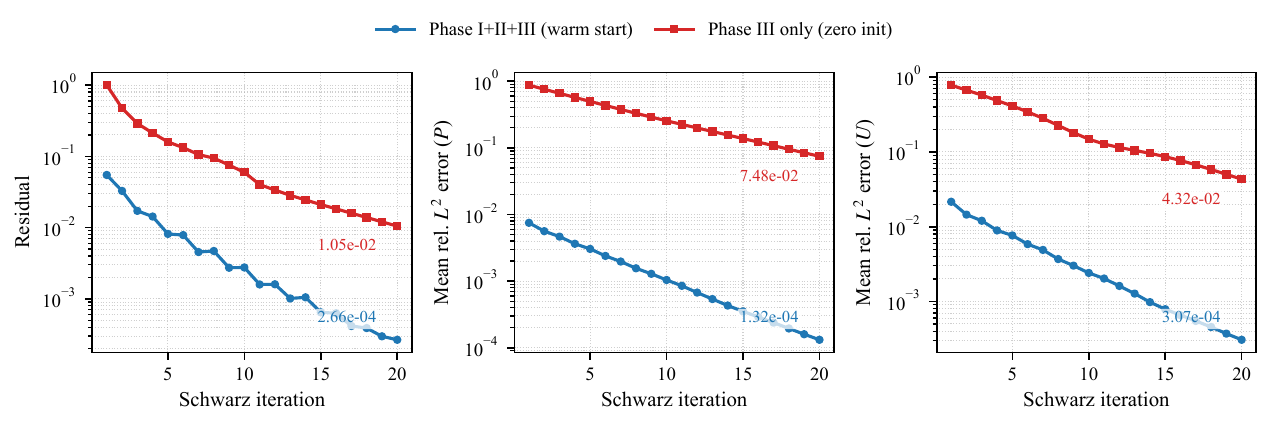}
    \caption{Warm-start ablation on the 7-segment arterial model. All curves show the SWR Numerical solver; Nektar1D is not plotted but serves as the ground-truth reference against which all errors are computed. Left: convergence residual. Middle and right: mean relative $L^2$ errors in pressure $P$ and velocity $U$. The warm-start strategy (Phase I+II+III, blue circles) initializes the SWR using the lumped RCR parameters (Phase I) for the initial simulation (Phase II), while the zero-initialization strategy (Phase III only, red squares) starts from zero interface data.}
    \label{fig:schwarz_initialization_comparison}
\end{figure}
Phases~I and~II of Algorithm~\ref{alg:main} aggregate lumped Windkessel parameters and perform a top-down warm-start solve before the iterative refinement begins in Phase~III. To isolate the effect of this initialization, we compare the full three-phase pipeline against a variant that skips Phases~I and~II and starts Phase~III from zero interface data. We use Nektar1D to independently solve each BU.

Fig.~\ref{fig:schwarz_initialization_comparison} shows the results on the 7-segment model using the SWR Numerical solver. The warm-start initialization begins with a residual roughly one order of magnitude lower than zero initialization and maintains this advantage throughout. We run $k=20$ iterations, which is sufficient to reveal the difference between the two strategies. After 20 Schwarz iterations, the complete Algorithm~\ref{alg:main} reaches a final residual of $2.66 \times 10^{-4}$, a mean pressure error of $1.32 \times 10^{-4}$, and a mean velocity error of $3.07 \times 10^{-4}$. In contrast, zero initialization converges to a residual of $1.05 \times 10^{-2}$, with pressure and velocity errors of $7.48 \times 10^{-2}$ and $4.32 \times 10^{-2}$, respectively, roughly two orders of magnitude worse.

The lumped-parameter warm-start is therefore not merely an acceleration trick but a key ingredient for SWR to reach an accurate solution within a practical number of iterations. A detailed analysis of SWR convergence, including the effects of network size, Windkessel parameters, vessel stiffness, heart rate, and inflow waveform, together with the segment-wise error distribution, is provided in Appendix~\ref{app:swr_convergence_size}.

\subsection{BU-Level Validation}

\begin{figure}
    \centering
    \includegraphics[width=\linewidth]{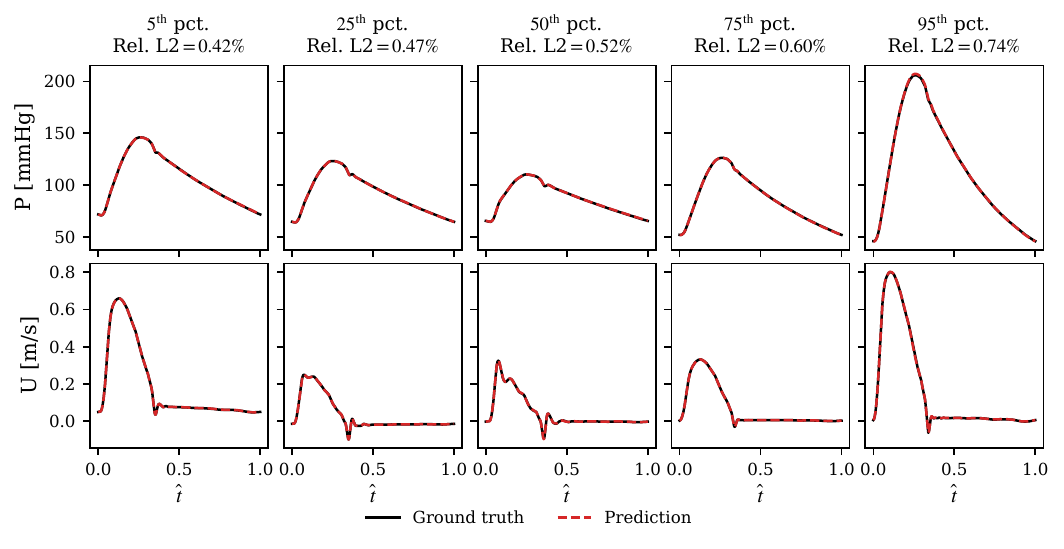}
    \vspace{-1.0pc}
    \caption{Pressure (top) and velocity (bottom) waveforms predicted by the BU operator (dashed red) versus the Nektar1D ground truth (black) for the 7-segment model. Each column corresponds to a test-set sample at the 5th, 25th, 50th, 75th, and 95th percentile of the relative $L^2$ error, with the error reported in the column header. The horizontal axis shows the normalized time $\hat{t} = t / T$. Results are shown at the midpoint history node of the parent segment. The prediction and the ground truth are qualitatively indistinguishable (red and black curves overlapping).}
    \label{fig:bu_percentile_curves_7seg}
\end{figure}

\begin{figure}
    \centering
    \includegraphics[width=\linewidth]{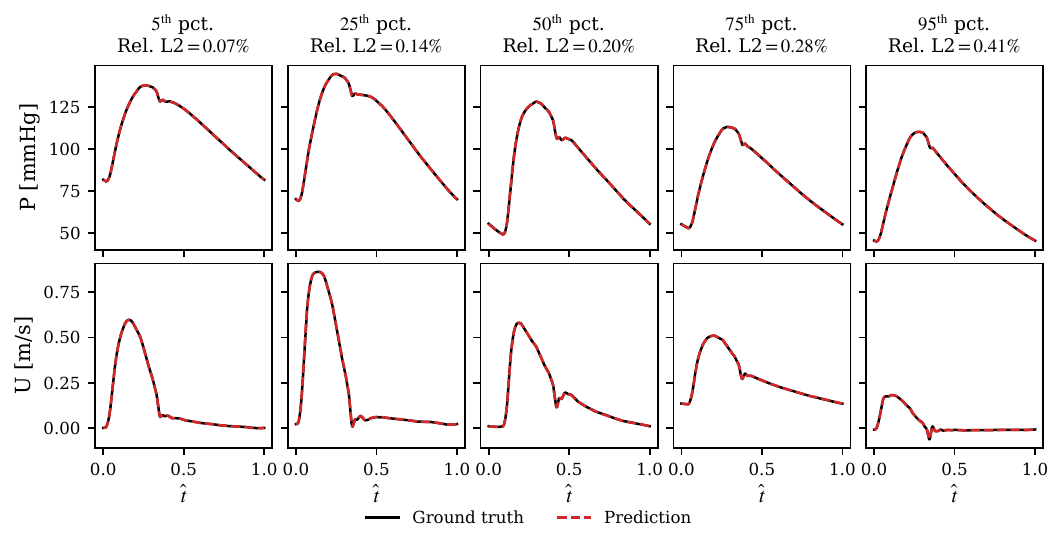}
    \vspace{-1.0pc}
    \caption{Same as Fig.~\ref{fig:bu_percentile_curves_7seg} but for the 55-segment full arterial tree model. The BU operator achieves a median relative $L^2$ error of 0.20\% across the test set, with 95th-percentile error below 0.41\%. The prediction and the ground truth are qualitatively indistinguishable.}
    \label{fig:bu_percentile_curves_55seg}
\end{figure}

Before deploying the trained BUFormer $\mathcal{G}_\theta$ as an operator inside the SWR loop, we first validate that it is accurate as a standalone local solver. All errors reported in this subsection are per-BU relative $L^2$ errors evaluated on the test set. Figures~\ref{fig:bu_percentile_curves_7seg} and~\ref{fig:bu_percentile_curves_55seg} show predicted versus ground-truth pressure and velocity waveforms at the 5th, 25th, 50th, 75th, and 95th percentiles of the test-set relative $L^2$ error distribution for the 7-segment and 55-segment models, respectively.

On the dataset for 7-segment model (Fig.~\ref{fig:bu_percentile_curves_7seg}), the predicted waveforms are visually indistinguishable from the ground truth across all percentiles. The median error is 0.52\%, and even the 95th-percentile case reaches only 0.74\%, indicating consistently low error with a tight distribution.

On the dataset for 55-segment model (Fig.~\ref{fig:bu_percentile_curves_55seg}), the operator remains accurate despite the greater diversity of BU configurations in the full arterial tree, with the median error at 0.20\% and the 95th-percentile error at 0.41\%. The predicted waveforms faithfully reproduce both the systolic peak and the dicrotic notch across the full range of test cases.

\begin{table}[t]
\footnotesize
\centering
\caption{Per-BC-type relative $L^2$ error of the BU operator on individual test-set units (no rollouts), mean $\pm$ std (\%). Type numbering follows Table~\ref{tab:bu_types}: BC type = (inlet)/(outlets) with $Q$ = prescribed flux, WK = Windkessel, $\mathbf{S}$ = full-state (prescribed flux and pressure); the two outlets are unordered. Type~1 is the root-unit configuration.}
\label{tab:rell2_per_bctype}
\begin{tabular}{llrcc}
\toprule
Model & BC type (inlet / outlets) & $n$ & \textbf{\(P\) rel \(L^2\)$(10^{-2})$} & \textbf{\(U\) rel \(L^2\)$(10^{-2})$} \\
\midrule
\textbf{7-segment} & all & 11529 & $0.701 \pm 0.234$ & $1.076 \pm 0.342$ \\
 & 1: $Q$ / (WK, WK) & 183 & $0.768 \pm 0.238$ & $0.950 \pm 0.384$ \\
 & 3: $Q$ / ($\mathbf{S}$, $\mathbf{S}$) & 3660 & $0.702 \pm 0.233$ & $0.995 \pm 0.349$ \\
 & 4: $\mathbf{S}$ / (WK, WK) & 7686 & $0.698 \pm 0.233$ & $1.117 \pm 0.329$ \\
\midrule
\textbf{55-segment} & all & 149823 & $0.219 \pm 0.124$ & $1.087 \pm 0.438$ \\
 & 1: $Q$ / (WK, WK) & 179 & $1.191 \pm 1.147$ & $1.753 \pm 0.323$ \\
 & 3: $Q$ / ($\mathbf{S}$, $\mathbf{S}$) & 5370 & $0.245 \pm 0.095$ & $1.676 \pm 0.244$ \\
 & 4: $\mathbf{S}$ / (WK, WK) & 42244 & $0.233 \pm 0.129$ & $0.782 \pm 0.437$ \\
 & 5: $\mathbf{S}$ / (WK, $\mathbf{S}$) & 75180 & $0.207 \pm 0.106$ & $1.126 \pm 0.357$ \\
 & 6: $\mathbf{S}$ / ($\mathbf{S}$, $\mathbf{S}$) & 26850 & $0.220 \pm 0.100$ & $1.337 \pm 0.341$ \\
\bottomrule
\end{tabular}
\end{table}

Table~\ref{tab:rell2_per_bctype} further evaluates the BU operator under different boundary condition configurations. Each BU is tested independently, allowing the influence of local boundary conditions on operator accuracy to be assessed. The results show consistently low relative $L^2$ errors across all BC types, including combinations of prescribed flow rate, Windkessel, and full-state $\mathbf{S}=(Q,P)$ conditions. The performance across different configurations indicates that the learned operator captures the local solution operator associated with diverse boundary conditions, providing the foundation for assembling accurate solutions over complex vascular networks.

These results confirm that the proposed BUFormer captures the local hemodynamic response with sufficient fidelity to serve as the prediction engine inside the decomposition solver. For reference, systematic comparisons between 1D and 3D hemodynamic simulations at arterial bifurcations report average relative errors below 2\%~\cite{Xiao2014}. The BU-level operator errors (median 0.20--0.52\%) are well within this margin, indicating that the learned operator introduces less discrepancy than the intrinsic gap between the 1D reduced-order and the 3D fully resolved models.

\subsection{Network-Level Performance}
\label{sec:network_performance}

\begin{table}[ht!]
\centering
\footnotesize
\caption{Speedup and agreement with the Nektar1D reference solution for the SWR operator and SWR numerical solvers, averaged over the test cases. Agreement is defined as $100\,(1 - \varepsilon)$\,\%, where $\varepsilon$ is the mean relative $L^2$ error. Nektar1D's average wall-clock times are 15.09\,s (7-seg) and 113.30\,s (55-seg).}
\label{tab:time_vs_rell2}
\begin{tabular}{lcc}
\toprule
 & SWR operator & SWR numerical \\
\midrule
\multicolumn{3}{l}{\textit{7-segment model}} \\
\quad Speedup vs.\ Nektar1D & $13.2\times$ & $0.04\times$ \\
\quad $P$ agreement (\%) & 99.68 & 99.98 \\
\quad $U$ agreement (\%) & 99.62 & 99.95 \\
\addlinespace
\multicolumn{3}{l}{\textit{55-segment model}} \\
\quad Speedup vs.\ Nektar1D & $17.3\times$ & $0.03\times$ \\
\quad $P$ agreement (\%) & 98.85 & 99.23 \\
\quad $U$ agreement (\%) & 97.79 & 98.28 \\
\bottomrule
\end{tabular}
\end{table}

We use the numerical SWR solver to prepare the dataset and train our BUFormer. The operator SWR solver then uses the trained BUFormer to replace the numerical solver for each BU, achieving a substantial speedup at the cost of a modest increase in error. We now report the tradeoff between speed and agreement with the Nektar1D reference among the SWR operator and SWR numerical solvers on the full vascular network. Table~\ref{tab:time_vs_rell2} reports speedups and agreement with the Nektar1D solution in pressure and velocity.
 
On the 7-segment model, the SWR operator solver achieves on average a $13.2\times$ speedup over Nektar1D with 99.68\% and 99.62\% agreement in pressure and velocity, respectively. On the 55-segment model, the speedup grows to $17.3\times$ with agreement above 97.7\% in both fields. These are per-case timings matched to Nektar1D's single-process CPU benchmark. GPU batching further reduces the amortized per-case cost, as analyzed in Appendix~\ref{app:parallel_timing}. The SWR numerical solver attains marginally higher agreement than the operator solver on both models (by roughly $0.3$--$0.5$ percentage points, see Table~\ref{tab:time_vs_rell2}) but is substantially slower, as each BU must be solved independently per iteration. However, this decomposition is what enables the operator training and replacement, thus the final speedup in the SWR operator solver. Moreover, reduced-order 1D Navier--Stokes models are known to deviate by 5--10\% from in-vivo or 3D reference solutions~\cite{Alastruey2011,Boileau2015,Xiao2014}. Our reported solvers achieve error levels below 3\%, well within this already-accepted range of model-form error. Our SWR operator solver therefore achieves a significant speedup while remaining viable.

\begin{figure}
    \centering
    \includegraphics[width=\linewidth]{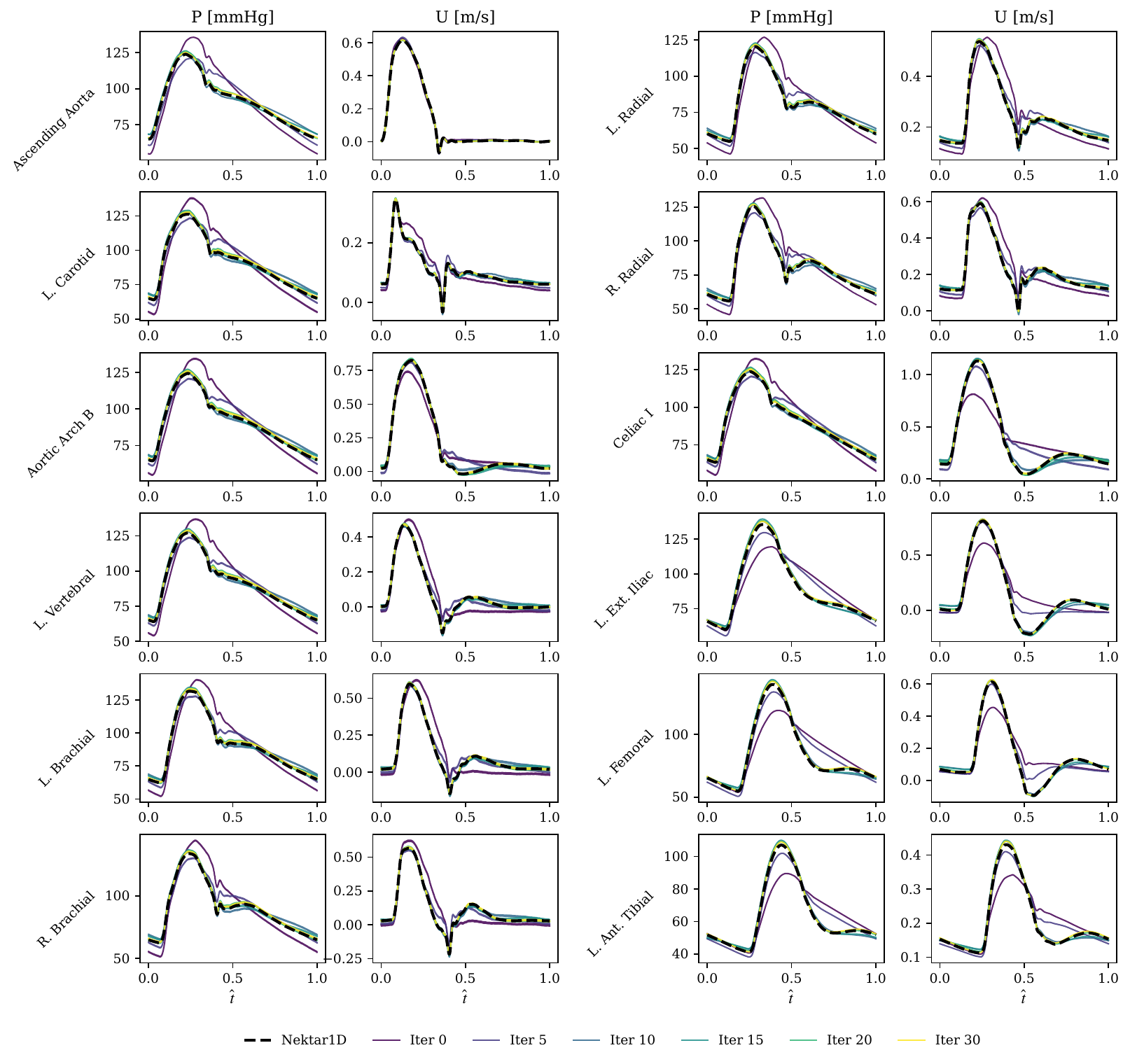}
    \vspace{-1.5pc}
    \caption{Convergence of the multiplicative SWR operator solution across iterations for the median-error test case. Each row corresponds to an anatomical segment spanning the arterial tree from the ascending aorta to the distal limb vessels. Pressure (left) and velocity (right) waveforms are shown at iterations 0, 5, 10, 15, 20, and 30, overlaid with the Nektar1D ground truth (black dashed). The horizontal axis shows the normalized time $\hat{t} = t / T$.}
    \label{fig:ddm_evolution_median_case}
\end{figure}

To illustrate how the waveform solutions evolve within a single case, Fig.~\ref{fig:ddm_evolution_median_case} shows pressure and velocity waveforms at selected segments on the 55-segment model across SWR iterations 0 through 30. At iteration 0 (the warm-start solution from Phase~II), the proximal segments already agree well with the Nektar1D ground truth, which confirms the results reported in \cite{epstein2015reducing}. After Phase II, the distal segments, particularly the L.\ Ext.\ Iliac, L.\ Femoral, and L.\ Ant.\ Tibial, show visible discrepancies in both amplitude and waveform shape. As iterations proceed, the solutions in all segments progressively align with the ground truth waveforms. By iteration 30 the remaining differences are negligible.

\begin{table}[ht]
\footnotesize
\centering
\caption{Interface pressure-jump error and mass-flux conservation residual at bifurcation-unit junctions, for the Nektar1D ground truth and the BU operator. Relative $L^2$ over time, mean $\pm$ std over the full test set (\%). The operator matches the ground-truth baseline, indicating the junction physics is preserved without any explicit continuity or conservation penalty in training.}
\label{tab:jump_conservation}
\begin{tabular}{lcccc}
\toprule
& \multicolumn{2}{c}{Pressure jump (\%)} & \multicolumn{2}{c}{Conservation residual (\%)} \\
\cmidrule(lr){2-3}\cmidrule(lr){4-5}
Model & Ground truth & BU operator & Ground truth & BU operator \\
\midrule
7-segment & $0.147 \pm 0.124$ & $0.151 \pm 0.119$ & $0.012 \pm 0.009$ & $0.320 \pm 0.148$ \\
55-segment & $0.350 \pm 0.393$ & $0.352 \pm 0.394$ & $0.022 \pm 0.056$ & $0.133 \pm 0.074$ \\
\bottomrule
\end{tabular}
\end{table}
We further examine the interface consistency of the converged BU solutions by measuring the pressure jump and mass-flux conservation residual at bifurcation junctions. At a bifurcation with parent $1$ and daughters $2,3$, pressure continuity requires $P_1=P_2=P_3$ and mass conservation requires $Q_1=Q_2+Q_3$. The pressure jump error is then the residual mismatch $|P_1-P_2|$ and $|P_1-P_3|$, and the mass-flux conservation error is the residual imbalance $|Q_1-Q_2-Q_3|$. These metrics quantify whether the locally predicted solutions from individual BUs satisfy the global coupling conditions after Schwarz waveform relaxation. As reported in Table~\ref{tab:jump_conservation}, the proposed BU operator achieves interface errors comparable to the Nektar1D reference across both vascular network sizes. The pressure-jump errors remain nearly unchanged relative to the ground-truth simulation, while the conservation residuals remain at a small level, indicating that the training objective effectively enforces junction consistency.

\subsection{Clinical Relevance Through PWV Robustness}

\begin{table}[t!]
\centering
\footnotesize
\caption{Comparison of pulse wave velocity (PWV) metrics on the 55-segment test set between the numerical solver Nektar1D and the SWR operator solver. The Nektar1D and SWR operator columns report the mean $\pm$ standard deviation over the test cases. The relative-error column reports the mean case-wise error, computed as $|PWV_\text{op} - PWV_\text{ref}| / PWV_\text{ref} \times 100\%$.}
\label{tab:pwv_comparison}
\begin{tabular}{lccc}
\toprule
PWV metric & Nektar1D (m/s) & SWR operator (m/s) & Rel.\ error (\%) \\
\midrule
aPWV (aortic, seg 1$\to$41) & $7.79 \pm 1.41$ & $7.72 \pm 1.37$ & $1.1$ \\
cfPWV (carotid--femoral) & $7.78 \pm 1.01$ & $7.76 \pm 1.00$ & $0.7$ \\
baPWV (brachial--ankle) & $8.34 \pm 0.96$ & $8.38 \pm 0.95$ & $0.7$ \\
faPWV (femoral--ankle) & $9.04 \pm 0.65$ & $9.07 \pm 0.63$ & $1.7$ \\
crPWV (carotid--radial) & $7.97 \pm 0.61$ & $7.90 \pm 0.64$ & $1.4$ \\
\bottomrule
\end{tabular}
\end{table}

While pointwise pressure and velocity errors are the standard numerical metrics for evaluating a solver, clinicians often care about derived biomarkers. Pulse wave velocity (PWV), computed from the transit time of the pressure pulse between two anatomical landmarks, is widely used to characterize arterial stiffness and vascular health~\cite{Willemet2015}. A solver that achieves low field errors but distorts the pulse propagation timing would be of limited clinical value.

Table~\ref{tab:pwv_comparison} compares five PWV metrics derived from the referencing Nektar1D solver and the SWR Operator solver on the 55-segment test set: aortic PWV (aPWV), carotid-femoral PWV (cfPWV), brachial-ankle PWV (baPWV), femoral-ankle PWV (faPWV), and carotid-radial PWV (crPWV). All five metrics are computed using the foot-to-foot method by estimating the pulse wave arrival-time difference between a proximal and a distal measurement site on the arterial tree. The pressure waveform feet is identified as the time of the diastolic pressure minimum preceding the first systolic peak. The PWV is then
\begin{equation}
  \mathrm{PWV} = \frac{\Delta L}{\Delta t},
  \label{eq:pwv_foot_to_foot}
\end{equation}
where $\Delta t = t_{\mathrm{foot}}^{\mathrm{dist}} - t_{\mathrm{foot}}^{\mathrm{prox}}$ is the foot arrival-time difference between the distal and proximal sites. The path length $\Delta L$ is computed as the direct vessel-path length along the connecting branch when the distal site is a descendant of the proximal site (aPWV), or as the difference in cumulative centerline distances from the aortic root to each site when the two sites lie on different branches of the arterial tree (cfPWV, baPWV, crPWV, faPWV). The specific measurement sites are: aPWV from the inlet of segment~1 (ascending aorta) to the outlet of segment~41 (abdominal aorta~E), cfPWV from the midpoint of segment~15 (L.\ common carotid) to the midpoint of segment~46 (L.\ femoral), baPWV from the midpoint of segment~21 (L.\ brachial) to the endpoint of segment~49 (L.\ anterior tibial), faPWV from the midpoint of segment~46 (L.\ femoral) to the midpoint of segment~49 (L.\ anterior tibial), and crPWV from the midpoint of segment~15 (L.\ common carotid) to the midpoint of segment~22 (L.\ radial).

Across all metrics, the operator-predicted values $PWV_\text{op}$ closely match the Nektar1D derived values $PWV_\text{ref}$; the mean relative errors range from 0.7\% (cfPWV, baPWV) to 1.7\% (faPWV), and the population-level distributions (mean $\pm$ std) are nearly indistinguishable between the two solvers. The aortic PWV (aPWV), the most clinically relevant indicator of central arterial stiffness, shows a relative error of only 1.1\%.

These results confirm that the low field-level errors reported in Section~\ref{sec:network_performance}  fall within clinically established tolerances for device measurement error. In particular, the operator-introduced PWV errors of 0.7--1.7\% are comfortably within the minimum clinically important difference established by the 2024 international consensus recommendations for PWV device validation~\cite{Spronck2024}.

\subsection{Generalization to Different Topologies}

\begin{figure*}[t!]
    \centering
    \includegraphics[width=0.78\textwidth]{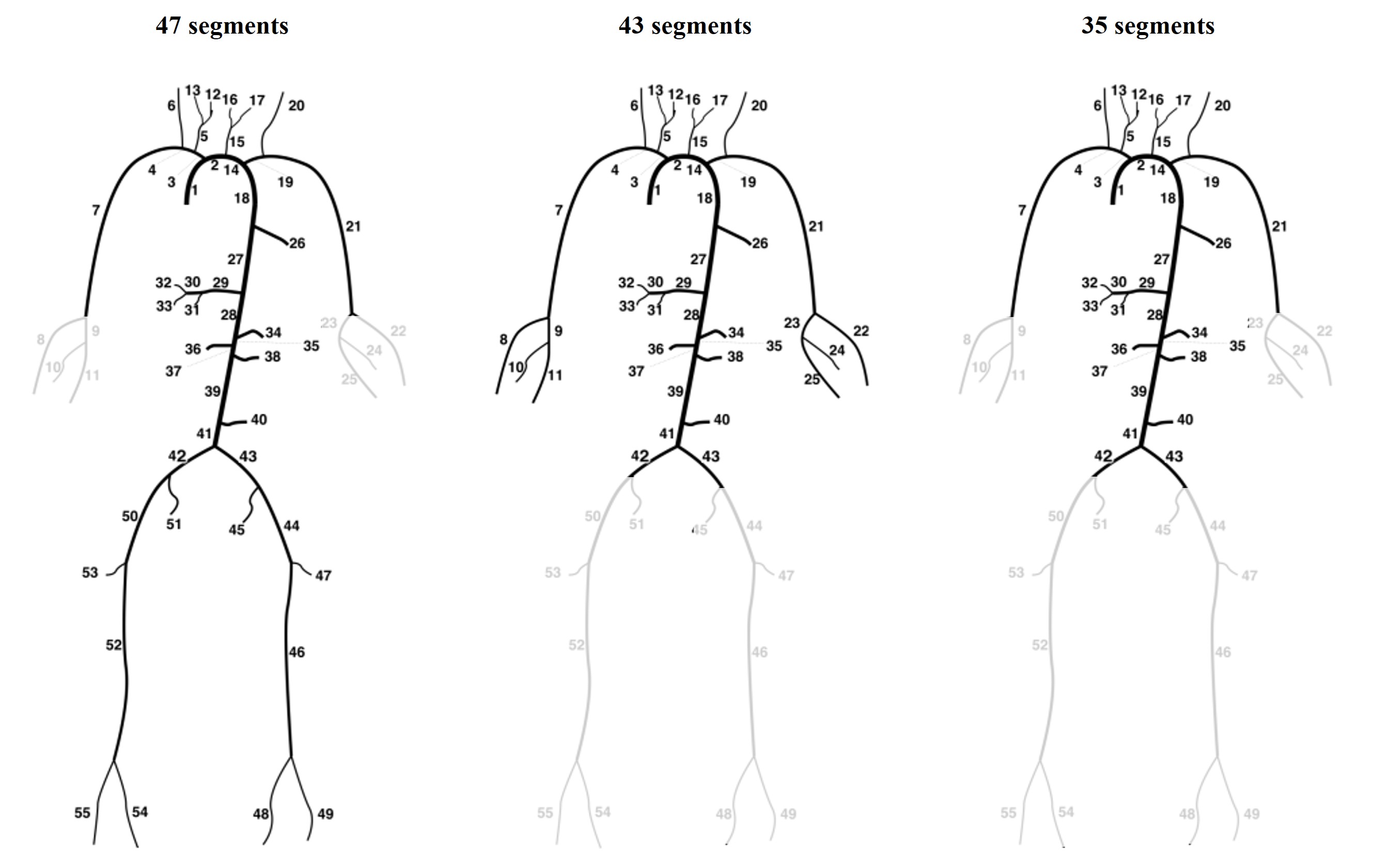}
    \caption{Pruned subsets of the 55-segment arterial tree used in the generalization experiment. Starting from the full network, a 47-segment network is obtained by removing the upper-limb descendants of segments~7 and~21. A 43-segment network is obtained by removing the lower-limb descendants of segments~42 and~43. Pruning all four yields the 35-segment network. The same operator checkpoint, trained on the full 55-segment tree, is applied to each pruned network without retraining.}
    \label{fig:subset_illustration}
\end{figure*}

\begin{table}[ht!]
\centering
\footnotesize
\caption{Subset-generalization evaluation averaged over the test cases for different target segment counts. Final relative \(L^2\) errors are the average error among all segments at the final SWR iteration (iteration 30).}
\begin{tabular}{c c c}
\hline
\textbf{Segments} & \textbf{\(P\) rel \(L^2\)$(10^{-2})$} & \textbf{\(U\) rel \(L^2\)$(10^{-2})$} \\
\hline
35 & 3.04 & 8.68 \\
43 & 1.72 & 5.83 \\
47 & 2.38 & 4.74 \\
55 (full) & 1.15 & 2.21 \\
\hline
\end{tabular}
\label{tab:subset_generalize_segments}
\end{table}

As noted in Remark~\ref{rmk:bc}, the BU operator $\mathcal{G}_{\theta}$ is conditioned only on unit-local parameters and does not depend on the global network topology. This means the same trained operator, using the checkpoint trained on the full 55-segment tree, can be applied without retraining to vascular networks with fewer segments. To test this, we prune the 55-segment tree by removing all descendants of selected segments, turning those segments into terminal vessels equipped with the Windkessel parameters obtained by lumping all their descendants via the procedure in Section~\ref{sec:rcr_lumping}. The 47-segment network removes the descendants of segments~7 (R.\ brachial) and~21 (L.\ brachial), the 43-segment network removes the descendants of segments~42 (L.\ common iliac) and~43 (R.\ common iliac), and the 35-segment network removes the descendants of all four  (Fig.~\ref{fig:subset_illustration}). We run Algorithm~\ref{alg:main} with the SWR Operator solver on each new network.

Table~\ref{tab:subset_generalize_segments} reports the final relative $L^2$ errors at iteration 30, averaged over the test cases. The errors increase moderately as the number of segments decreases: from 1.15\% (pressure) and 2.21\% (velocity) on the full 55-segment tree to 3.04\% and 8.68\% on the most aggressively pruned 35-segment network. Even in the most aggressive case, these errors remain within the 5--10\% tolerance generally accepted in one-dimensional hemodynamic modeling~\cite{Reymond2009,Xiao2014,Alastruey2011}. This graceful degradation is a direct consequence of the local operator design, the operator never sees the global tree structure, so removing segments does not invalidate the learned mapping.

This property is practically relevant in clinical settings where imaging data are incomplete or certain vessel segments fall below imaging resolution. The proposed framework remains applicable without modification, offering empirical robustness that purely global operators cannot match without retraining.

\subsection{OOD with Elastic Vascular Aging and Hypertension}
\label{sec:ood_operator}

We test the learned operator on the fixed 55-segment tree under clinically plausible but out-of-distribution conditions. Concretely, we consider hypertensive patients in their 40s to 50s whose arteries have stiffened prematurely, as in Early Vascular Aging (EVA): elevated
elastic and muscular stiffness (up to $2.5\times$ and $1.4\times$ baseline) combined with the younger, near-normal vessel diameters of that age, and reference pressure raised to the hypertensive end ($1.4\times$). Since the training data couples the two along a healthy aging trajectory, this decoupling of high stiffness from small diameter is the source of the distribution shift. Heart rate, inflow, and peripheral resistance stay within nominal ranges. We run operator SWR for 30 iterations, reporting in Table~\ref{tab:ood_operator} the mean relative $L^2$ error in pressure $P$ and velocity $U$ against the ground truth.

\begin{table}[ht!]
\centering
\footnotesize
\caption{Operator SWR under out-of-distribution elastic vascular aging and hypertension, on the
fixed 55-segment tree. Final relative \(L^2\) errors at the final SWR iteration (iteration 30),
reported as mean \(\pm\) std across the 16 cases in each group. All cases converge.}
\begin{tabular}{c c c}
\hline
\textbf{Group} & \textbf{\(P\) rel \(L^2\)$(10^{-2})$} & \textbf{\(U\) rel \(L^2\)$(10^{-2})$} \\
\hline
40\,yr body & $6.2 \pm 2.4$ & $8.9 \pm 3.6$ \\
50\,yr body & $9.0 \pm 3.2$ & $15.0 \pm 5.0$ \\
All & $7.6 \pm 3.2$ & $12.0 \pm 5.3$ \\
\hline
\end{tabular}
\label{tab:ood_operator}
\end{table}

All cases remain stable in the tested scenarios, with a moderate increase in prediction error as the physiological conditions become more challenging: the mean $P$/$U$ errors increase from $6.2\%$/$8.9\%$ for the milder 40\,yr phenotype to $9.0\%$/$15.0\%$ for the stiffer 50\,yr phenotype. Similar to in-distribution cases, errors may accumulate over Schwarz iterations and along the vascular tree, making the most extrapolated conditions and distal segments more challenging to approximate. These results suggest that the proposed operator maintains robust performance across the tested aging and hypertension-related parameter variations.

\subsection{Discussion}

The experimental results support four main findings. First, the initial simulation with lumped RCR parameters (Phases~I and~II) is essential. Removing Phases~I and~II degrades SWR accuracies roughly two orders of magnitude (Fig.~\ref{fig:schwarz_initialization_comparison}). Second, the BU operator $\mathcal{G}_\theta$ is accurate as a standalone local solver, with median relative $L^2$ errors of 0.52\% on the 7-segment model and 0.20\% on the 55-segment model (Figures~\ref{fig:bu_percentile_curves_7seg}--\ref{fig:bu_percentile_curves_55seg}). Third, the operator-driven SWR solver achieves $13\times$ and $17\times$ speedups over the Nektar1D solver on the 7-segment and 55-segment networks, respectively, with agreement of 97.8--99.7\% (Table~\ref{tab:time_vs_rell2}), and the resulting PWV biomarkers agree with the Nektar1D reference to within 1--2\% (Table~\ref{tab:pwv_comparison}). Fourth, the locality of the BU operator design confers robustness to incomplete vascular topologies. The same checkpoint trained on the 55-segment tree only performs moderately worse on pruned arterial trees (Table~\ref{tab:subset_generalize_segments}), without any retraining.

It is worth noting that the role of the domain decomposition in our method is not to accelerate the numerical solution directly, but to make the problem amenable to surrogate modeling. It partitions the network into small, structurally uniform BUs, each of which can be solved by a single neural operator evaluation, and it allows these local evaluations to be carried out in parallel across the network at each SWR iteration. Together with the GPU parallelism provided by widely available deep learning frameworks, this yields a substantial speedup.

The proposed BU-based framework is independent of the specific neural operator backbone. Although we employ a transformer-based operator as the main implementation due to its strong representation capability, the proposed decomposition and Schwarz coupling strategy can also be combined with alternative operator-learning architectures. We verify this flexibility by replacing the transformer backbone with a Fourier Neural Operator (FNO) and reporting the corresponding results in Appendix~\ref{app:backbone}. In addition, we conduct self-ablation studies with different transformer configurations to examine the sensitivity of the method to architectural choices like the number of Fourier modes and the balanced sampling mechanism. The detailed results are provided in Appendix~\ref{app:ablation}.
\section{Conclusion}
\label{sec:conclusion}

We have presented an operator learning framework for fast hemodynamic simulation in vascular networks. The method decomposes the network into BUs, aggregates downstream effects via Windkessel parameter lumping, and replaces the numerical solver at each BU with a trained BU operator, implemented as the BUFormer architecture. A three-phase algorithm (bottom-up parameter aggregation, top-down warm-start initialization, and operator-driven multiplicative Schwarz waveform relaxation (SWR)) assembles the local predictions into a globally consistent solution.

Experiments on the 7-segment and 55-segment arterial tree models demonstrate that the approach achieves significant wall-clock speedups over conventional 1D numerical simulation with mean relative $L^2$ errors of order 1\% in both pressure and velocity. The BU operator is accurate as a standalone local solver (median error below 0.5\%), and the resulting full-network solutions preserve clinically relevant pulse wave velocity biomarkers to within 1--2\% of the Nektar1D reference. The locality of the operator design further enables generalization to vascular networks with different tree-like topologies, including incomplete networks with missing segments, without retraining.

Several future opportunities emerge from this work. While the BU decomposition and the Schwarz waveform relaxation apply to general 1D solvers, this study validates the framework under specific healthy-subject physiological assumptions. A natural next step is to extend the training dataset to pathological conditions such as severe stenoses and aneurysms, which introduce distinct hemodynamic regimes beyond the healthy physiology modeled here. Our framework may also reduce the cost of patient-specific model calibration, where physiological parameters are inferred from clinical measurements via repeated forward queries~\cite{kissas2020machine}. Combined with patient-specific data, such capabilities could support surgical pre-planning scenarios such as virtual stent placement~\cite{zhao2021patient}. Besides, the basic unit concept can be extended beyond the simple bifurcation to encompass any local topology that admits a tractable reduced-order description, such as trifurcations or side-branch junctions, further broadening the class of vascular networks the framework can represent. Equipping the training stage with a physics-informed loss may also contribute to improved training efficiency and inference performance. 
Finally, we acknowledge that a rigorous convergence analysis of the multiplicative SWR scheme for 1D Navier--Stokes vessel networks remains an open problem, though a plausible direction is apparent. The 1D blood flow model can be recast as a telegrapher (damped-wave) equation by linearizing about the diastolic state~\cite{tamburrelli2010spectral}. For this equation, the convergence of overlapping Schwarz waveform relaxation has been analyzed~\cite{alkhaleel2024telegrapher}. These results, however, concern a single segment without branching, whereas a complete proof must also establish well-posedness and convergence on the coupled network with bifurcations considered here. Such well-posedness and convergence have been shown for simplified cases, such as a three-segment binary tree under Robin boundary conditions~\cite{maity2022existence}, and reconciling the local subproblems under our full-state $(Q,P)$ coupling with these results remains the key open step. We plan to combine these directions into a complete theoretical analysis of the scheme in future work. 

\section*{Acknowledgements}
This work was partially supported by the Office of Naval Research (ONR), Army Research Office (ARO), the National Science Foundation (NSF), the Simons Foundation, RGC-GRF project 11309925, Mathematical Formalization of GIS and gifts or awards from Google, Amazon, and Apple. The funders had no role in the study design; in the collection, analysis, or interpretation of data; in the writing of the manuscript; or in the decision to submit the article for publication. Thanks to the InnoHK project at Hong Kong Centre for Cerebro-cardiovascular Health Engineering (COCHE). The authors thank Dr.\ Jordi Alastruey for sharing the source code of Nektar1D.

\bibliographystyle{siamplain}
\bibliography{references}

@article{crosetto2011parallel,
  title={{Parallel Algorithms for Fluid-Structure Interaction Problems in Haemodynamics}},
  author={Crosetto, Paolo and Deparis, Simone and Fourestey, Gilles and Quarteroni, Alfio},
  journal={{SIAM Journal on Scientific Computing}},
  volume={33},
  number={4},
  pages={1598--1622},
  year={2011},
  doi={ 10.1137/090772836 },
  publisher={SIAM}
}

@article{quarteroni2004mathematical,
  title={{Mathematical Modelling and Numerical Simulation of the Cardiovascular System}},
  author={Quarteroni, Alfio and Formaggia, Luca},
  journal={{Handbook of Numerical Analysis}},
  volume={12},
  pages={3--127},
  year={2004},
  doi={ 10.1016/s1570-8659(03)12001-7 },
  publisher={Elsevier}
}

@article{raissi2019physics,
  title={{Physics-Informed Neural Networks: A Deep Learning Framework for Solving Forward and Inverse Problems Involving Nonlinear Partial Differential Equations}},
  author={Raissi, Maziar and Perdikaris, Paris and Karniadakis, George E},
  journal={{Journal of Computational Physics}},
  volume={378},
  pages={686--707},
  year={2019},
  doi={ 10.1016/j.jcp.2018.10.045 },
  publisher={Elsevier}
}

@article{yan2022impact,
  title={{Impact of Pressure Wire on Fractional Flow Reserve and Hemodynamics of the Coronary Arteries: A Computational and Clinical Study}},
  author={Yan, Zhengzheng and Yao, Zhifeng and Guo, Weifeng and Shang, Dandan and Chen, Rongliang and Liu, Jia and Cai, Xiao-Chuan and Ge, Junbo},
  journal={{IEEE Transactions on Biomedical Engineering}},
  volume={70},
  number={5},
  pages={1683--1691},
  year={2023},
  doi={10.1109/TBME.2022.3225188},
  publisher={IEEE}
}

@article{bazilevs2008isogeometric,
  title={{Isogeometric Fluid-Structure Interaction: Theory, Algorithms, and Computations}},
  author={Bazilevs, Yuri and Calo, Victor M and Hughes, Thomas JR and Zhang, Yongjie},
  journal={{Computational Mechanics}},
  volume={43},
  pages={3--37},
  year={2008},
  doi={ 10.1007/s00466-008-0315-x },
  publisher={Springer}
}

@article{figueroa2006coupled,
  title={{A Coupled Momentum Method for Modeling Blood Flow in Three-Dimensional Deformable Arteries}},
  author={Figueroa, C Alberto and Vignon-Clementel, Irene E and Jansen, Kenneth E and Hughes, Thomas JR and Taylor, Charles A},
  journal={{Computer Methods in Applied Mechanics and Engineering}},
  volume={195},
  number={41-43},
  pages={5685--5706},
  year={2006},
  doi={ 10.1016/j.cma.2005.11.011 },
  publisher={Elsevier}
}

@article{pijls1996measurement,
  title={{Measurement of Fractional Flow Reserve to Assess the Functional Severity of Coronary-Artery Stenoses}},
  author={Pijls, Nico HJ and de Bruyne, Bernard and Peels, Kathinka and van der Voort, Pepijn H and Bonnier, Hans JRM and Bartunek, Jozef and Koolen, Jacques J},
  journal={{New England Journal of Medicine}},
  volume={334},
  number={26},
  pages={1703--1708},
  year={1996},
  doi={ 10.1056/nejm199606273342604 },
  publisher={Mass Medical Soc}
}

@article{bukavc2013fluid,
  title={{Fluid-Structure Interaction in Blood Flow Capturing Non-Zero Longitudinal Structure Displacement}},
  author={Buka{\v{c}}, Martina and {\v{C}}ani{\'c}, Sun{\v{c}}ica and Glowinski, Roland and Tamba{\v{c}}a, Josip and Quaini, Annalisa},
  journal={{Journal of Computational Physics}},
  volume={235},
  pages={515--541},
  year={2013},
  doi={ 10.1016/j.jcp.2012.08.033 },
  publisher={Elsevier}
}

@article{wang2018higher,
  title={{A Higher-Order Discontinuous Galerkin/Arbitrary Lagrangian Eulerian Partitioned Approach to Solving Fluid--Structure Interaction Problems with Incompressible, Viscous Fluids and Elastic Structures}},
  author={Wang, Yifan and Quaini, Annalisa and {\v{C}}ani{\'c}, Sun{\v{c}}ica},
  journal={{Journal of Scientific Computing}},
  volume={76},
  pages={481--520},
  year={2018},
  doi={ 10.1007/s10915-017-0629-y },
  publisher={Springer}
}

@article{peskin1972flow,
  title={{Flow Patterns Around Heart Valves: A Numerical Method}},
  journal={{Journal of Computational Physics}},
  volume={10},
  number={2},
  pages={252-271},
  year={1972},
  issn={0021-9991},
  doi={https://doi.org/10.1016/0021-9991(72)90065-4},
  author={Charles S Peskin}
}

@article{formaggia2001coupling,
  title={{On the Coupling of 3D and 1D Navier--Stokes Equations for Flow Problems in Compliant Vessels}},
  author={Formaggia, Luca and Gerbeau, Jean-Fr{\'e}d{\'e}ric and Nobile, Fabio and Quarteroni, Alfio},
  journal={{Computer Methods in Applied Mechanics and Engineering}},
  volume={191},
  number={6-7},
  pages={561--582},
  year={2001},
  doi={ 10.1016/s0045-7825(01)00302-4 },
  publisher={Elsevier}
}

@article{zhang2024meshless,
  title={{A Meshless Solver for Blood Flow Simulations in Elastic Vessels Using a Physics-Informed Neural Network}},
  author={Zhang, Han and Chan, Raymond H and Tai, Xue-Cheng},
  journal={{SIAM Journal on Scientific Computing}},
  volume={46},
  number={4},
  pages={C479--C507},
  year={2024},
  doi={ 10.1137/23m1622696 },
  publisher={SIAM}
}

@article{zhang2024learning,
  title={{A Learning-Based Framework for Topology-Preserving Segmentation Using Quasiconformal Mappings}},
  author={Zhang, Han and Lui, Lok Ming},
  journal={{Neurocomputing}},
  volume={600},
  pages={128124},
  year={2024},
  doi={ 10.1016/j.neucom.2024.128124 },
  publisher={Elsevier}
}

@inproceedings{li2020fourier,
  title={{Fourier Neural Operator for Parametric Partial Differential Equations}},
  author={Li, Zongyi and Kovachki, Nikola and Azizzadenesheli, Kamyar and Liu, Burigede and Bhattacharya, Kaushik and Stuart, Andrew and Anandkumar, Anima},
  booktitle={{International Conference on Learning Representations}},
  year={2021},
  url={https://openreview.net/forum?id=c8P9NQVtmnO}
}

@article{lu2021learning,
  title={{Learning Nonlinear Operators via DeepONet Based on the Universal Approximation Theorem of Operators}},
  author={Lu, Lu and Jin, Pengzhan and Pang, Guofei and Zhang, Zhongqiang and Karniadakis, George Em},
  journal={{Nature Machine Intelligence}},
  volume={3},
  number={3},
  pages={218--229},
  year={2021},
  doi={ 10.1038/s42256-021-00302-5 },
  publisher={Nature Publishing Group UK London}
}

@article{maity2022existence,
  title={Existence and uniqueness of maximal strong solution of a 1D blood flow in a network of vessels},
  author={Maity, Debayan and Raymond, Jean-Pierre and Roy, Arnab},
  journal={Nonlinear Analysis: Real World Applications},
  volume={63},
  pages={103405},
  year={2022},
  publisher={Elsevier}
}

@article{pegolotti2024learning,
  title={{Learning Reduced-Order Models for Cardiovascular Simulations with Graph Neural Networks}},
  author={Pegolotti, Luca and Pfaller, Martin R and Rubio, Natalia L and Ding, Ke and Brufau, Rita Brugarolas and Darve, Eric and Marsden, Alison L},
  journal={{Computers in Biology and Medicine}},
  volume={168},
  pages={107676},
  year={2024},
  doi={ 10.1016/j.compbiomed.2023.107676 },
  publisher={Elsevier}
}

@article{velikorodny2025deep,
  title={{Deep Operator Learning for Blood Flow Modelling in Stenosed Vessels}},
  author={Velikorodny, Alexis and Lu, Lu and Dudenkov, Vladimir and Glanz, Victor and Chernyavsky, Boris and Neylon, Antoinette and Smits, Pieter C},
  journal={{NPJ Artificial Intelligence}},
  volume={1},
  number={1},
  pages={35},
  year={2025},
  doi={10.1038/s44387-025-00035-5},
  publisher={Nature Publishing Group UK London}
}

@article{chan1994domain,
  title={{Domain Decomposition Algorithms}},
  author={Chan, Tony F and Mathew, Tarek P},
  journal={{Acta Numerica}},
  volume={3},
  pages={61--143},
  year={1994},
  doi={ 10.1017/s0962492900002427 },
  publisher={Cambridge University Press}
}

@article{epstein2015reducing,
    author = {Epstein, Sally and Willemet, Marie and Chowienczyk, Phil J. and Alastruey, Jordi},
    title = {{Reducing the Number of Parameters in 1D Arterial Blood Flow Modeling: Less Is More for Patient-Specific Simulations}},
    journal = {{American Journal of Physiology-Heart and Circulatory Physiology}},
    volume = {309},
    number = {1},
    pages = {H222-H234},
    year = {2015},
  doi={ 10.1152/ajpheart.00857.2014 },
}

@article{perdikaris2014fractional,
  title={{Fractional-Order Viscoelasticity in One-Dimensional Blood Flow Models}},
  author={Perdikaris, Paris and Karniadakis, George Em},
  journal={{Annals of Biomedical Engineering}},
  volume={42},
  number={5},
  pages={1012--1023},
  year={2014},
  doi={ 10.1007/s10439-014-0970-3 },
  publisher={Springer}
}

@article{formaggia2013physical,
  title={{On the Physical Consistency Between Three-Dimensional and One-Dimensional Models in Haemodynamics}},
  author={Formaggia, Luca and Quarteroni, Alfio and Vergara, Christian},
  journal={{Journal of Computational Physics}},
  volume={244},
  pages={97--112},
  year={2013},
  doi={ 10.1016/j.jcp.2012.08.001 },
  publisher={Elsevier}
}

@inproceedings{formaggia2002one,
  title={{A One Dimensional Model for Blood Flow: Application to Vascular Prosthesis}},
  author={Formaggia, Luca and Nobile, Fabio and Quarteroni, Alfio},
  booktitle={{Mathematical Modeling and Numerical Simulation in Continuum Mechanics}},
  pages={137--153},
  year={2002},
  doi={ 10.1007/978-3-642-56288-4_10 },
  organization={Springer}
}

@article{formaggia2003one,
  title={{One-Dimensional Models for Blood Flow in Arteries}},
  author={Formaggia, Luca and Lamponi, Daniele and Quarteroni, Alfio},
  journal={{Journal of Engineering Mathematics}},
  volume={47},
  number={3},
  pages={251--276},
  year={2003},
  doi={ 10.1023/b:engi.0000007980.01347.29 },
  publisher={Springer}
}

@article{grinberg2011modeling,
  title={{Modeling Blood Flow Circulation in Intracranial Arterial Networks: A Comparative 3D/1D Simulation Study}},
  author={Grinberg, Leopold and Cheever, Elizabeth and Anor, Tomer and Madsen, Joseph R and Karniadakis, GE},
  journal={{Annals of Biomedical Engineering}},
  volume={39},
  number={1},
  pages={297--309},
  year={2011},
  doi={ 10.1007/s10439-010-0132-1 },
  publisher={Springer}
}

@article{zhang2025parametrized,
  title={{Parametrized Sampling for 3D Blood Simulation in Deformable Vessels Using Physics-Informed Neural Networks}},
  author={Zhang, Han and Li, Lingfeng and Tai, Xue-Cheng and Chan, Raymond Hon-Fu},
  journal={{Journal of Computational and Applied Mathematics}},
  pages={117197},
  volume={477},
  year={2026},
  doi={ 10.1016/j.cam.2025.117197 },
  publisher={Elsevier}
}

@article{cheng2025surrogate,
  title={{Surrogate Modeling of Heat Transfer Under Flow Fluctuation Conditions Using Fourier Basis-Deep Operator Network with Uncertainty Quantification}},
  author={Cheng, Qiyun and Sahadath, Md Hossain and Yang, Huihua and Pan, Shaowu and Ji, Wei},
  journal={{Progress in Nuclear Energy}},
  volume={188},
  pages={105895},
  year={2025},
  doi={ 10.1016/j.pnucene.2025.105895 },
  publisher={Elsevier}
}

@article{abueidda2025deepokan,
  title={{Deepokan: Deep Operator Network Based on Kolmogorov Arnold Networks for Mechanics Problems}},
  author={Abueidda, Diab W and Pantidis, Panos and Mobasher, Mostafa E},
  journal={{Computer Methods in Applied Mechanics and Engineering}},
  volume={436},
  pages={117699},
  year={2025},
  doi={ 10.1016/j.cma.2024.117699 },
  publisher={Elsevier}
}

@article{rezaei2025digitalizing,
  title={{Digitalizing Metallic Materials from Image Segmentation to Multiscale Solutions via Physics Informed Operator Learning}},
  author={Rezaei, Shahed and Taghikhani, Kianoosh and Viardin, Alexandre and Asl, Reza Najian and Harandi, Ali and Jagtap, Nikhil Vijay and Bailly, David and Naber, Hannah and Gramlich, Alexander and Brepols, Tim and others},
  journal={{NPJ Computational Materials}},
  volume={11},
  number={1},
  pages={262},
  year={2025},
  doi={10.1038/s41524-025-01718-y},
  publisher={Nature Publishing Group UK London}
}

@article{koric2023data,
  title={{Data-Driven and Physics-Informed Deep Learning Operators for Solution of Heat Conduction Equation with Parametric Heat Source}},
  author={Koric, Seid and Abueidda, Diab W},
  journal={{International Journal of Heat and Mass Transfer}},
  volume={203},
  pages={123809},
  year={2023},
  doi={ 10.1016/j.ijheatmasstransfer.2022.123809 },
  publisher={Elsevier}
}

@article{cai2021deepm,
  title={{DeepM\&Mnet: Inferring the Electroconvection Multiphysics Fields Based on Operator Approximation by Neural Networks}},
  author={Cai, Shengze and Wang, Zhicheng and Lu, Lu and Zaki, Tamer A and Karniadakis, George Em},
  journal={{Journal of Computational Physics}},
  volume={436},
  pages={110296},
  year={2021},
  doi={ 10.1016/j.jcp.2021.110296 },
  publisher={Elsevier}
}

@inproceedings{wang2024cvit,
  title={{CViT: Continuous Vision Transformer for Operator Learning}},
  author={Wang, Sifan and Seidman, Jacob H and Sankaran, Shyam and Wang, Hanwen and Pappas, George J and Perdikaris, Paris},
  booktitle={{International Conference on Learning Representations}},
  year={2025},
  url={https://openreview.net/forum?id=cRnCcuLvyr}
}

@article{wang2021learning,
  title={{Learning the Solution Operator of Parametric Partial Differential Equations with Physics-Informed DeepONets}},
  author={Wang, Sifan and Wang, Hanwen and Perdikaris, Paris},
  journal={{Science Advances}},
  volume={7},
  number={40},
  pages={eabi8605},
  year={2021},
  doi={ 10.1126/sciadv.abi8605 },
  publisher={American Association for the Advancement of Science}
}

@incollection{goswami2023physics,
  title={{Physics-Informed Deep Neural Operator Networks}},
  author={Goswami, Somdatta and Bora, Aniruddha and Yu, Yue and Karniadakis, George Em},
  booktitle={{Machine Learning in Modeling and Simulation. Computational Methods in Engineering \& the Sciences}},
  pages={219--254},
  year={2023},
  doi={ 10.1007/978-3-031-36644-4_6 },
  publisher={Springer}
}

@incollection{Alastruey2012,
  author    = {Alastruey, Jordi and Parker, Kim H. and Sherwin, Spencer J.},
  title     = {{Arterial Pulse Wave Haemodynamics}},
  booktitle = {{11th International Conference on Pressure Surges}},
  pages     = {401--443},
  year      = {2012},
  publisher = {BHR Group},
  url       = {https://kclpure.kcl.ac.uk/portal/en/publications/arterial-pulse-wave-haemodynamics/}
}

@article{rueth2021quasi,
  author  = {R\"{u}th, Benjamin and Uekermann, Benjamin and Mehl, Miriam and Birken, Philipp and Monge, Azahar and Bungartz, Hans-Joachim},
  title   = {{Quasi-{Newton} Waveform Iteration for Partitioned Surface-Coupled Multiphysics Applications}},
  journal = {{International Journal for Numerical Methods in Engineering}},
  year    = {2021},
  volume  = {122},
  number  = {19},
  pages   = {5236--5257},
  doi     = {10.1002/nme.6443}
}

@article{dong2024convergence,
  author  = {Dong, W. B. and Tang, H. S.},
  title   = {{Convergence Analysis of {Schwarz} Waveform Relaxation Method to Compute Coupled Advection--Diffusion--Reaction Equations}},
  journal = {{Mathematics and Computers in Simulation}},
  year    = {2024},
  volume  = {218},
  pages   = {462--481},
  doi     = {10.1016/j.matcom.2023.11.026}
}

@article{lemarie2010optimized,
  author  = {Lemari\'{e}, Florian and Debreu, Laurent and Blayo, Eric},
  title   = {{Optimized {Schwarz} Waveform Relaxation for the Primitive Equations of the Ocean}},
  journal = {{SIAM Journal on Scientific Computing}},
  year    = {2010},
  volume  = {32},
  number  = {5},
  pages   = {2908--2936},
  doi     = {10.1137/090770059}
}

@book{quarteroni1999domain,
  author    = {Quarteroni, Alfio and Valli, Alberto},
  title     = {{Domain Decomposition Methods for Partial Differential Equations}},
  publisher = {Oxford University Press},
  year      = {1999},
  doi={ 10.1093/oso/9780198501787.001.0001 },
}

@article{gander1998space,
  author  = {Gander, Martin J. and Stuart, Andrew M.},
  title   = {{Space-Time Continuous Analysis of Waveform Relaxation for the Heat Equation}},
  journal = {{SIAM Journal on Scientific Computing}},
  year    = {1998},
  volume  = {19},
  number  = {6},
  pages   = {2014--2031},
  doi     = {10.1137/S1064827596305337}
}

@article{gander2003optimal,
  author  = {Gander, Martin J. and Halpern, Laurence and Nataf, Fr{\'e}d{\'e}ric},
  title   = {{Optimal {Schwarz} Waveform Relaxation for the One Dimensional Wave Equation}},
  journal = {{SIAM Journal on Numerical Analysis}},
  year    = {2003},
  volume  = {41},
  number  = {5},
  pages   = {1643--1681},
  doi     = {10.1137/S003614290139559X}
}

@article{gander2007optimized,
  author  = {Gander, Martin J. and Halpern, Laurence},
  title   = {{Optimized {Schwarz} Waveform Relaxation Methods for Advection Reaction Diffusion Problems}},
  journal = {{SIAM Journal on Numerical Analysis}},
  year    = {2007},
  volume  = {45},
  number  = {2},
  pages   = {666--697},
  doi     = {10.1137/050642137}
}

@article{caetano2010schwarz,
  title={{Schwarz Waveform Relaxation Algorithms for Semilinear Reaction-Diffusion Equations}},
  author={Caetano, Filipa and Gander, Martin J and Halpern, Laurence and Szeftel, J{\'e}r{\'e}mie},
  journal={{Networks and Heterogeneous Media}},
  volume={5},
  number={3},
  pages={487--505},
  year={2010},
  doi={ 10.3934/nhm.2010.5.487 },
  publisher={American Institute of Mathematical Sciences}
}

@inproceedings{descombes2011schwarz,
  title={{Schwarz Waveform Relaxation Methods for Systems of Semi-Linear Reaction-Diffusion Equations}},
  author={Descombes, St{\'e}phane and Dolean, Victorita and Gander, Martin J},
  booktitle={{Domain Decomposition Methods in Science and Engineering XIX}},
  pages={151--158},
  year={2011},
  doi={ 10.1007/978-3-642-11304-8_49 },
  publisher={Springer}
}

@article{Reymond2009,
  author  = {Reymond, Philippe and Merenda, Fabrice and Perren, Fabienne and R\"{u}fenacht, Daniel and Stergiopulos, Nikos},
  title   = {{Validation of a One-Dimensional Model of the Systemic Arterial Tree}},
  journal = {{American Journal of Physiology--Heart and Circulatory Physiology}},
  year    = {2009},
  volume  = {297},
  pages   = {H208--H222},
  doi     = {10.1152/ajpheart.00037.2009}
}

@article{Willemet2015,
  author  = {Willemet, Marie and Chowienczyk, Phil and Alastruey, Jordi},
  title   = {{A Database of Virtual Healthy Subjects to Assess the Accuracy of Foot-to-Foot Pulse Wave Velocities for Estimation of Aortic Stiffness}},
  journal = {{American Journal of Physiology--Heart and Circulatory Physiology}},
  year    = {2015},
  volume  = {309},
  number  = {4},
  pages   = {H663--H675},
  doi     = {10.1152/ajpheart.00175.2015}
}

@article{Alastruey2011,
  author  = {Alastruey, Jordi and Khir, Ashraf W. and Matthys, Koen S. and Segers, Patrick and Sherwin, Spencer J. and Verdonck, Pascal R. and Parker, Kim H. and Peir\'{o}, Joaquim},
  title   = {{Pulse Wave Propagation in a Model Human Arterial Network: Assessment of 1-D Visco-Elastic Simulations Against In Vitro Measurements}},
  journal = {{Journal of Biomechanics}},
  year    = {2011},
  volume  = {44},
  number  = {12},
  pages   = {2250--2258},
  doi     = {10.1016/j.jbiomech.2011.05.041}
}

@article{Boileau2015,
  author  = {Boileau, Etienne and Nithiarasu, Perumal and Blanco, Pablo J. and M\"{u}ller, Lucas O. and Fossan, Fredrik Eikeland and Hellevik, Leif Rune and Donders, Wouter P. and Huberts, Wouter and Willemet, Marie and Alastruey, Jordi},
  title   = {{A Benchmark Study of Numerical Schemes for One-Dimensional Arterial Blood Flow Modelling}},
  journal = {{International Journal for Numerical Methods in Biomedical Engineering}},
  year    = {2015},
  volume  = {31},
  number  = {10},
  pages   = {e02732},
  doi     = {10.1002/cnm.2732}
}

@article{Xiao2014,
  author  = {Xiao, Nan and Alastruey, Jordi and Figueroa, C. Alberto},
  title   = {{A Systematic Comparison Between 1-D and 3-D Hemodynamics in Compliant Arterial Models}},
  journal = {{International Journal for Numerical Methods in Biomedical Engineering}},
  year    = {2014},
  volume  = {30},
  number  = {2},
  pages   = {204--231},
  doi     = {10.1002/cnm.2598}
}

@article{westerhof1971artificial,
  author  = {Westerhof, N. and Elzinga, G. and Sipkema, P.},
  title   = {{An Artificial Arterial System for Pumping Hearts}},
  journal = {{Journal of Applied Physiology}},
  volume  = {31},
  number  = {5},
  pages   = {776--781},
  year    = {1971},
  doi     = {10.1152/jappl.1971.31.5.776}
}

@article{westerhof2009arterial,
  author  = {Westerhof, N. and Lankhaar, J.-W. and Westerhof, B. E.},
  title   = {{The Arterial Windkessel}},
  journal = {{Medical \& Biological Engineering \& Computing}},
  volume  = {47},
  number  = {2},
  pages   = {131--141},
  year    = {2009},
  doi     = {10.1007/s11517-008-0359-2}
}

@article{fleeter2020multilevel,
  author  = {Fleeter, Casey M. and Geraci, Gianluca and Schiavazzi, Daniele E. and Kahn, Andrew M. and Marsden, Alison L.},
  title   = {{Multilevel and Multifidelity Uncertainty Quantification for Cardiovascular Hemodynamics}},
  journal = {{Computer Methods in Applied Mechanics and Engineering}},
  volume  = {365},
  pages   = {113030},
  year    = {2020},
  doi     = {10.1016/j.cma.2020.113030}
}

@article{brault2017uncertainty,
  author  = {Brault, Antoine and Dumas, Laurent and Lucor, Didier},
  title   = {{Uncertainty Quantification of Inflow Boundary Condition and Proximal Arterial Stiffness---Coupled Effect on Pulse Wave Propagation in a Vascular Network}},
  journal = {{International Journal for Numerical Methods in Biomedical Engineering}},
  volume  = {33},
  number  = {10},
  pages   = {e2859},
  year    = {2017},
  doi     = {10.1002/cnm.2859}
}

@article{charlton2019modeling,
  author  = {Charlton, Peter H. and Mariscal Harana, Jorge and Vennin, Samuel and Li, Ye and Chowienczyk, Phil and Alastruey, Jordi},
  title   = {{Modeling Arterial Pulse Waves in Healthy Aging: a Database for in Silico Evaluation of Hemodynamics and Pulse Wave Indexes}},
  journal = {{American Journal of Physiology -- Heart and Circulatory Physiology}},
  volume  = {317},
  number  = {5},
  pages   = {H1062--H1085},
  year    = {2019},
  doi     = {10.1152/ajpheart.00218.2019}
}

@article{kissas2020machine,
  author  = {Kissas, Georgios and Yang, Yibo and Hwuang, Eileen and Witschey, Walter R. and Detre, John A. and Perdikaris, Paris},
  title   = {{Machine Learning in Cardiovascular Flows Modeling: Predicting Arterial Blood Pressure from Non-invasive 4D Flow MRI Data Using Physics-Informed Neural Networks}},
  journal = {{Computer Methods in Applied Mechanics and Engineering}},
  volume  = {358},
  pages   = {112623},
  year    = {2020},
  doi     = {10.1016/j.cma.2019.112623}
}

@article{fossan2021machine,
  author  = {Fossan, Fredrik E. and M{\"u}ller, Lucas O. and Sturdy, Jacob and Braaten, Anders T. and J{\o}rgensen, Arve and Wiseth, Rune and Hellevik, Leif R.},
  title   = {{Machine Learning Augmented Reduced-Order Models for FFR-Prediction}},
  journal = {{Computer Methods in Applied Mechanics and Engineering}},
  volume  = {384},
  pages   = {113892},
  year    = {2021},
  doi     = {10.1016/j.cma.2021.113892}
}

@article{Spronck2024,
  author  = {Spronck, Bart and Terentes-Printzios, Dimitrios and Avolio, Alberto P. and Boutouyrie, Pierre and Guala, Andrea and Jeron\v{c}i\'{c}, Ana and Laurent, St\'{e}phane and others},
  title   = {{2024 Recommendations for Validation of Noninvasive Arterial Pulse Wave Velocity Measurement Devices}},
  journal = {{Hypertension}},
  year    = {2024},
  volume  = {81},
  number  = {1},
  pages   = {183--192},
  doi     = {10.1161/HYPERTENSIONAHA.123.21618}
}

@article{zhao2021patient,
  author  = {Zhao, Shijia and Wu, Wei and Samant, Saurabhi and Khan, Behram and Kassab, Ghassan S. and Watanabe, Yusuke and Murasato, Yoshinobu and Sharzehee, Mohammadali and Makadia, Janaki and Zolty, Daniel and Panagopoulos, Anastasios and Burzotta, Francesco and Migliavacca, Francesco and Johnson, Thomas W. and Lefevre, Thierry and Lassen, Jens Flensted and Brilakis, Emmanouil S. and Bhatt, Deepak L. and Dangas, George and Chiastra, Claudio and Stankovic, Goran and Louvard, Yves and Chatzizisis, Yiannis S.},
  title   = {Patient-specific Computational Simulation of Coronary Artery Bifurcation Stenting},
  journal = {Scientific Reports},
  year    = {2021},
  volume  = {11},
  pages   = {16486},
  doi     = {10.1038/s41598-021-95026-2}
}

@article{alkhaleel2024telegrapher,
  author  = {Al-Khaleel, Mohammad D. and Gander, Martin J. and Kumbhar, Pratik M.},
  title   = {{Optimized {Schwarz} Waveform Relaxation Methods for the Telegrapher Equation}},
  journal = {{SIAM Journal on Scientific Computing}},
  year    = {2024},
  volume  = {46},
  number  = {6},
  pages   = {A3528--A3551},
  doi     = {10.1137/24M1642962}
}

@inproceedings{tamburrelli2010spectral,
  author    = {Tamburrelli, Vincenzopio and Ferranti, Francesco and Antonini, Giulio and Cristina, Saverio and Dhaene, Tom and Knockaert, Luc},
  title     = {{Spectral models for 1D blood flow simulations}},
  booktitle = {{2010 Annual International Conference of the IEEE Engineering in Medicine and Biology Society (EMBC)}},
  year      = {2010},
  pages     = {2598--2601},
  doi       = {10.1109/IEMBS.2010.5626619}
}

\newpage
\appendix
\section{Hyperparameters}
\label{app:hyperparams}

Table~\ref{tab:hyperparams} lists the architecture and training hyperparameters used for the two experimental settings. The 55-segment model uses a larger embedding dimension and more transformer layers to accommodate the greater diversity of bifurcation unit configurations in the full arterial tree. All experiments are conducted on a single NVIDIA A100 GPU.

\begin{table}[ht!]
\centering
\footnotesize
\caption{Hyperparameters of the BUFormer operator for the two experimental settings. FFN denotes the feedforward network within each Transformer layer, and FP16 denotes half-precision floating-point training.}
\begin{tabular}{l c c}
\hline
\textbf{Hyperparameter} & \textbf{7-Segment} & \textbf{55-Segment} \\
\hline
\multicolumn{3}{l}{\textit{Architecture}} \\
Embedding dimension $h$          & 512   & 1024  \\
Fourier modes $K$                & 128   & 128    \\
Transformer layers $N$           & 8     & 12    \\
Attention heads                  & 8     & 8     \\
FFN hidden dimension             & 2048  & 4096  \\
Spatial history points $N_x$     & 10    & 10    \\
Temporal resolution $N_t$        & 1024  & 1024  \\
\hline
\multicolumn{3}{l}{\textit{Training}} \\
Optimizer                        & \multicolumn{2}{c}{AdamW} \\
Peak learning rate               & $3 \times 10^{-4}$ & $3 \times 10^{-4}$ \\
Minimum learning rate            & $1 \times 10^{-6}$ & $1 \times 10^{-6}$ \\
LR schedule                      & \multicolumn{2}{c}{Cosine annealing} \\
Weight decay                     & $1 \times 10^{-4}$ & $1 \times 10^{-4}$ \\
Gradient clipping norm           & 1.0   & 1.0   \\
Batch size                       & 1024  & 800   \\
Epochs                           & 200   & 80    \\
BC type balancing                & Yes   & Yes   \\
Mixed precision (FP16)           & Yes   & Yes   \\
\hline
\end{tabular}
\label{tab:hyperparams}
\end{table}

\section{Baseline Arterial Parameters}
\label{app:baseline_params}

Table~\ref{tab:55segments} lists the baseline geometric, material, and boundary parameters used for all simulations. $L$ is the segment length, $D_{\mathrm{in}} \rightarrow D_{\mathrm{out}}$ are the inlet and outlet diameters (assumed to vary linearly), and $a$ is the pulse wave velocity (PWV) parameter used to compute the wall stiffness $\beta$ via~\eqref{eq:beta_from_a}. Leaf segments have prescribed total peripheral resistance $R$ and compliance $C$, while internal segments bifurcate into child segments. All arterial segments are assumed to be purely elastic. Simulations use outflow pressure $P_\text{inf}=1.33 \text{ kPa (10 mmHg)}$, blood density $\rho=1050 \text{ kg/m}^3$, dynamic viscosity $\mu=4.0 \text{ mPa} \cdot \text{s}$, and momentum correction coefficient $\alpha=1.1$. These physical values are standard assumptions following previous studies~\cite{Willemet2015}, and realistic values exhibit only minor deviations. Our framework can handle different values in principle, but capturing the minor deviations here is not necessary for the scope of this work.

\begin{table*}
\centering
\scriptsize
\caption{Baseline parameters of the 55-segment arterial tree model~\cite{Willemet2015}. Segments marked with $\dagger$ form the 7-segment subset.}
\begin{tabular}{r l c c c c c}
\hline
\textbf{Seg} & \textbf{Arterial Segment} & \makecell{\textbf{Length},\\ ($L$) \textbf{cm}} & \makecell{\textbf{Diameter}\\ ($D_{\mathrm{in}} \rightarrow D_{\mathrm{out}}$) \textbf{mm}} & \makecell{\textbf{PWV}\\ $a$} & \makecell{\textbf{Peripheral} \\ \textbf{Resistance} $R$ \\ $10^{10}$ Pa$\cdot$s$\cdot$m$^{-3}$} & \makecell{\textbf{Peripheral} \\ \textbf{Compliance} $C$ \\ $10^{-10}$ m$^3$/Pa} \\
\hline
 1$^\dagger$ & Ascending aorta          &  5.80 & $30.27 \rightarrow 30.17$ & 11.0 &    -- &    -- \\
 2$^\dagger$ & Aortic arch A            &  2.30 & $25.84 \rightarrow 24.71$ & 11.0 &    -- &    -- \\
 3$^\dagger$ & Brachiocephalic          &  3.90 & $20.80 \rightarrow 18.54$ & 11.0 &    -- &    -- \\
 4$^\dagger$ & R.\ subclavian           &  3.90 & $11.84 \rightarrow 9.27$  & 11.0 &    -- &    -- \\
 5$^\dagger$ & R.\ common carotid       & 10.80 & $11.29 \rightarrow 5.86$  & 11.0 &    -- &    -- \\
 6           & R.\ vertebral            & 17.10 & $3.81 \rightarrow 2.88$   & 12.0 & 0.451 & 0.903 \\
 7           & R.\ brachial             & 48.50 & $8.34 \rightarrow 4.84$   & 12.0 &    -- &    -- \\
 8           & R.\ radial               & 27.00 & $3.81 \rightarrow 3.19$   & 12.0 & 0.396 & 0.988 \\
 9           & R.\ ulnar A              &  7.70 & $3.81 \rightarrow 3.50$   & 12.0 &    -- &    -- \\
10           & R.\ interosseous         &  9.10 & $2.16 \rightarrow 1.85$   & 12.0 & 6.323 & 0.325 \\
11           & R.\ ulnar B              & 19.70 & $3.29 \rightarrow 2.88$   & 12.0 & 0.396 & 0.770 \\
12           & R.\ internal carotid     & 20.50 & $5.87 \rightarrow 4.43$   & 12.0 & 0.188 & 2.585 \\
13           & R.\ external carotid     & 18.70 & $2.57 \rightarrow 1.54$   & 12.0 & 1.043 & 1.928 \\
14$^\dagger$ & Aortic arch B            &  4.50 & $22.03 \rightarrow 21.41$ & 11.0 &    -- &    -- \\
15$^\dagger$ & L.\ common carotid       & 16.00 & $10.04 \rightarrow 5.02$  & 11.0 &    -- &    -- \\
16           & L.\ internal carotid     & 20.50 & $4.43 \rightarrow 3.43$   & 12.0 & 0.188 & 1.888 \\
17           & L.\ external carotid     & 18.70 & $1.97 \rightarrow 1.25$   & 12.0 & 1.043 & 1.730 \\
18           & Thoracic aorta A         &  6.00 & $20.59 \rightarrow 19.45$ & 11.0 &    -- &    -- \\
19           & L.\ subclavian           &  3.90 & $11.33 \rightarrow 8.76$  & 11.0 &    -- &    -- \\
20           & L.\ vertebral            & 17.00 & $3.81 \rightarrow 2.88$   & 12.0 & 0.451 & 0.903 \\
21           & L.\ brachial             & 48.50 & $8.34 \rightarrow 4.84$   & 12.0 &    -- &    -- \\
22           & L.\ radial               & 27.00 & $3.60 \rightarrow 2.88$   & 12.0 & 0.396 & 0.848 \\
23           & L.\ ulnar A              &  7.70 & $4.43 \rightarrow 4.43$   & 12.0 &    -- &    -- \\
24           & L.\ interosseous         &  9.10 & $1.85 \rightarrow 1.85$   & 12.0 & 6.323 & 0.278 \\
25           & L.\ ulnar B              & 19.70 & $4.22 \rightarrow 3.81$   & 12.0 & 0.396 & 1.300 \\
26           & Intercostals             &  9.20 & $12.97 \rightarrow 9.78$  & 11.0 & 0.600 & 10.433 \\
27           & Thoracic aorta B         & 12.00 & $16.99 \rightarrow 13.27$ & 11.0 &    -- &    -- \\
28           & Abdominal aorta A        &  6.10 & $12.56 \rightarrow 12.56$ & 11.0 &    -- &    -- \\
29           & Celiac A                 &  2.30 & $8.03 \rightarrow 7.10$   & 11.0 &    -- &    -- \\
30           & Celiac B                 &  2.30 & $5.35 \rightarrow 5.04$   & 11.0 &    -- &    -- \\
31           & Hepatic                  &  7.60 & $5.56 \rightarrow 4.53$   & 12.0 & 0.272 & 2.053 \\
32           & Gastric                  &  8.20 & $3.29 \rightarrow 3.09$   & 12.0 & 0.406 & 0.820 \\
33           & Splenic                  &  7.20 & $4.32 \rightarrow 4.02$   & 12.0 & 0.174 & 1.400 \\
34           & Superior mesenteric      &  6.80 & $8.13 \rightarrow 7.31$   & 11.0 & 0.070 & 4.805 \\
35           & Abdominal aorta B        &  2.30 & $11.84 \rightarrow 11.63$ & 11.0 &    -- &    -- \\
36           & L.\ renal                &  3.70 & $5.35 \rightarrow 5.35$   & 11.0 & 0.085 & 2.310 \\
37           & Abdominal aorta C        &  2.30 & $12.15 \rightarrow 12.15$ & 11.0 &    -- &    -- \\
38           & R.\ renal                &  3.70 & $5.35 \rightarrow 5.35$   & 11.0 & 0.085 & 2.310 \\
39           & Abdominal aorta D        & 12.20 & $11.94 \rightarrow 11.32$ & 11.0 &    -- &    -- \\
40           & Inferior mesenteric      &  5.80 & $4.84 \rightarrow 3.28$   & 12.0 & 0.516 & 1.333 \\
41           & Abdominal aorta E        &  2.30 & $11.12 \rightarrow 10.71$ & 11.0 &    -- &    -- \\
42           & L.\ common iliac         &  6.80 & $8.13 \rightarrow 7.20$   & 11.0 &    -- &    -- \\
43           & R.\ common iliac         &  6.80 & $8.13 \rightarrow 7.20$   & 11.0 &    -- &    -- \\
44           & L.\ external iliac       & 16.60 & $6.59 \rightarrow 6.28$   & 11.0 &    -- &    -- \\
45           & L.\ internal iliac       &  5.80 & $4.12 \rightarrow 4.12$   & 12.0 & 0.596 & 1.368 \\
46           & L.\ femoral              & 50.90 & $5.35 \rightarrow 3.91$   & 12.0 &    -- &    -- \\
47           & L.\ deep femoral         & 14.50 & $4.12 \rightarrow 3.81$   & 12.0 & 0.358 & 1.268 \\
48           & L.\ posterior tibial     & 36.90 & $3.19 \rightarrow 2.88$   & 12.0 & 0.358 & 0.743 \\
49           & L.\ anterior tibial      & 39.80 & $2.68 \rightarrow 2.37$   & 12.0 & 0.419 & 0.513 \\
50           & R.\ external iliac       & 16.60 & $6.59 \rightarrow 6.28$   & 11.0 &    -- &    -- \\
51           & R.\ internal iliac       &  5.80 & $4.12 \rightarrow 4.12$   & 12.0 & 0.596 & 1.368 \\
52           & R.\ femoral              & 50.90 & $5.35 \rightarrow 3.91$   & 12.0 &    -- &    -- \\
53           & R.\ deep femoral         & 14.50 & $4.12 \rightarrow 3.81$   & 12.0 & 0.358 & 1.268 \\
54           & R.\ posterior tibial     & 36.90 & $3.19 \rightarrow 2.88$   & 12.0 & 0.358 & 0.743 \\
55           & R.\ anterior tibial      & 39.80 & $2.68 \rightarrow 2.37$   & 12.0 & 0.419 & 0.513 \\
\hline
\end{tabular}
\label{tab:55segments}
\end{table*}

\section{Training Data Normalization}
\label{app:normalization}
The operator inputs and targets span several orders of magnitude and carry different units (pressures $\sim 10^4$~Pa, velocities $\sim 1$~m/s, resistances $\sim 10^{8}$, compliances $\sim 10^{-9}$), so every field is normalized before training. All statistics are computed once on the training split, frozen, and reused for validation, testing, and inference. Predictions are mapped back to physical units for the SWR coupling and error reporting.

Each waveform is mapped to $\hat{t} = t/T \in [0,1]$, giving all samples identical temporal support regardless of cardiac period or solver time step. The predicted pressure $P$ and velocity $U$ are each standardized by a single global mean and standard deviation taken over all training samples, history points, and time steps. The network operates in this normalized space; the inverse maps recover physical units, and the relative $L^2$ accuracy is evaluated per field after denormalization. For boundary conditions, inlet flow $Q$ is standardized by its own statistics, while pressure or velocity boundary waveforms reuse the field statistics.

The operator's scalar inputs comprise segment diameters, lengths, wall stiffness, reference pressure, cardiac period, and lumped Windkessel boundary conditions. Each is directly normalized by their train set statistics. Quantities that span many orders of magnitude are log-transformed before normalization.

\section{SWR Convergence Analysis}

We examine how the SWR convergence behavior depends on different factors including network size on the anatomical tree (\ref{app:swr_convergence_size}), total impedance and compliance (\ref{app:swr_convergence_RC}), vessel stiffness (\ref{app:swr_convergence_stiffness}), heart rate (\ref{app:swr_convergence_heartrate}), and inflow waveform shapes (\ref{app:swr_convergence_inflow}). Across all these factors, multiplicative SWR converges consistently within the ranges considered. Few breakdowns occur only at non-physiological extremes and stem from the numerical solver, not the SWR coupling. All experiments here use the \emph{numerical} Nektar1D solver at each bifurcation unit, so the convergence reported is a property of the SWR coupling itself.

\subsection{Convergence vs. network size}
\label{app:swr_convergence_size}

\begin{figure}[ht!]
    \centering
    \includegraphics[width=\linewidth]{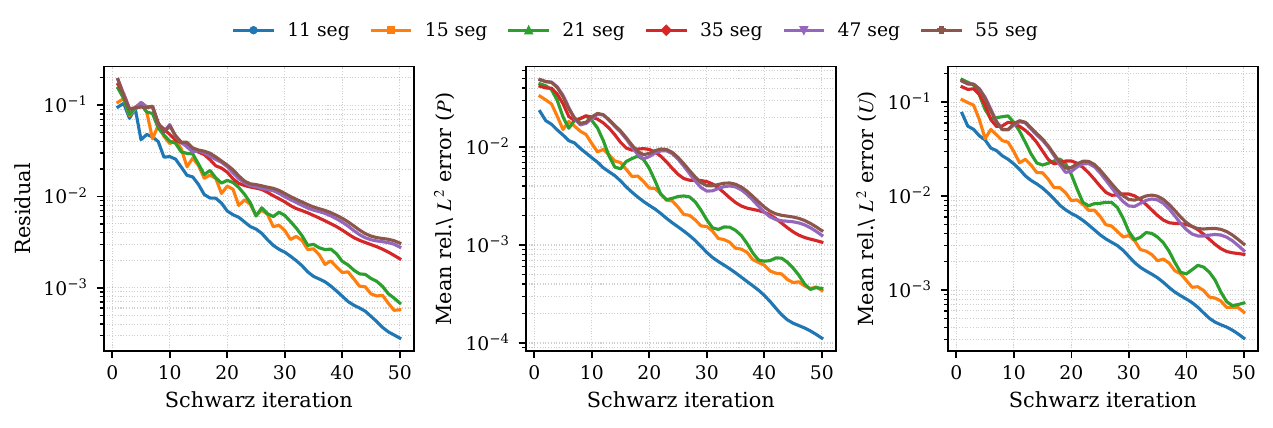}
    \caption{Numerical Schwarz waveform relaxation (SWR) convergence on arterial networks of 11, 15, 21, 35, 47, and 55 segments, all using the warm-start initialization (Phases~I+II+III of Algorithm~\ref{alg:main}). Left: convergence residual $\|\mathbf{S}^{(k)} - \mathbf{S}^{(k-1)}\|$ versus iteration. Middle and right: mean relative $L^2$ error in pressure $P$ and velocity $U$ against the Nektar1D ground truth.}
    \label{fig:convergence_vs_network_size}
\end{figure}

\begin{figure}[ht!]
    \centering
    \includegraphics[width=\linewidth]{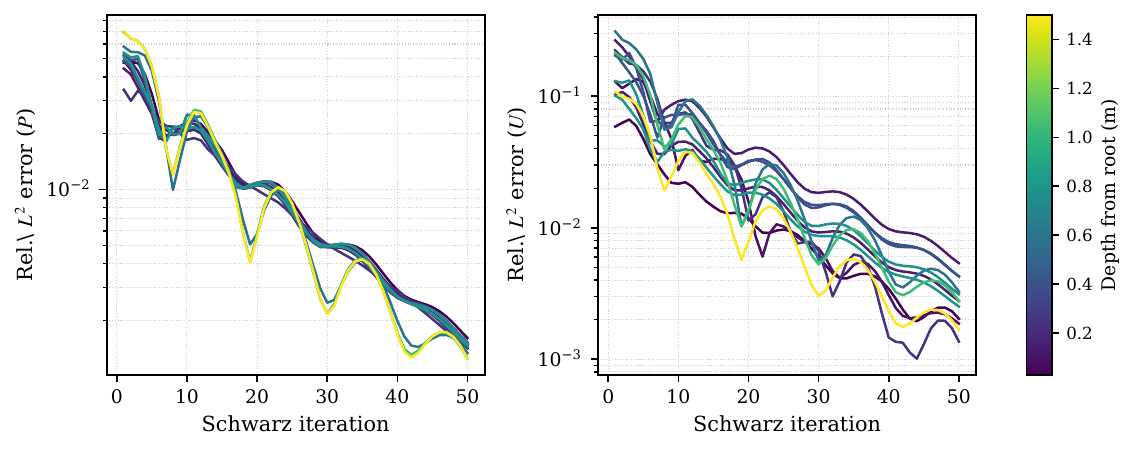}
    \caption{Per-segment relative $L^2$ error in pressure $P$ (left) and velocity $U$ (right) versus Schwarz iteration on the 55-segment arterial model, for the same 12 anatomical segments shown in Fig.~\ref{fig:ddm_evolution_median_case}. Each curve is colored by the cumulative centerline distance from the aortic root to the segment midpoint.}
    \label{fig:per_segment_error_convergence}
\end{figure}

We examine how the SWR convergence rate scales with network size and how errors distribute across individual segments.

Fig.~\ref{fig:convergence_vs_network_size} shows the convergence residual and mean relative $L^2$ errors for six network sizes ranging from 11 to 55 segments, all run with the warm-start initialization for 50 Schwarz iterations. All networks converge monotonically in the mean, with smaller networks converging faster. The 11-segment network reaches a residual of $2.81 \times 10^{-4}$ and mean pressure and velocity errors of $1.12 \times 10^{-4}$ and $3.08 \times 10^{-4}$, respectively, by iteration 50. The full 55-segment network converges more slowly, reaching a residual of $3.09 \times 10^{-3}$ with mean pressure and velocity errors of $1.40 \times 10^{-3}$ and $3.08 \times 10^{-3}$. The gap between the smallest and largest networks is roughly one order of magnitude in all three metrics, indicating that convergence degrades gracefully with network size.

Fig.~\ref{fig:per_segment_error_convergence} shows the per-segment error trajectories on the 55-segment model for the same 12 anatomical segments depicted in Fig.~\ref{fig:ddm_evolution_median_case}. Proximal segments near the aortic root converge fastest, while distal segments such as the L.\ femoral and L.\ anterior tibial lag behind. This pattern reflects the top-down structure of the multiplicative Schwarz sweep; since information from the root propagates outward through successive iterations, deeper segments require more iterations to receive accurate boundary data from their ancestors.

\subsection{Network topology}
\label{app:swr_convergence_topology}

\begin{figure}[ht!]
    \centering
    \includegraphics[width=\linewidth]{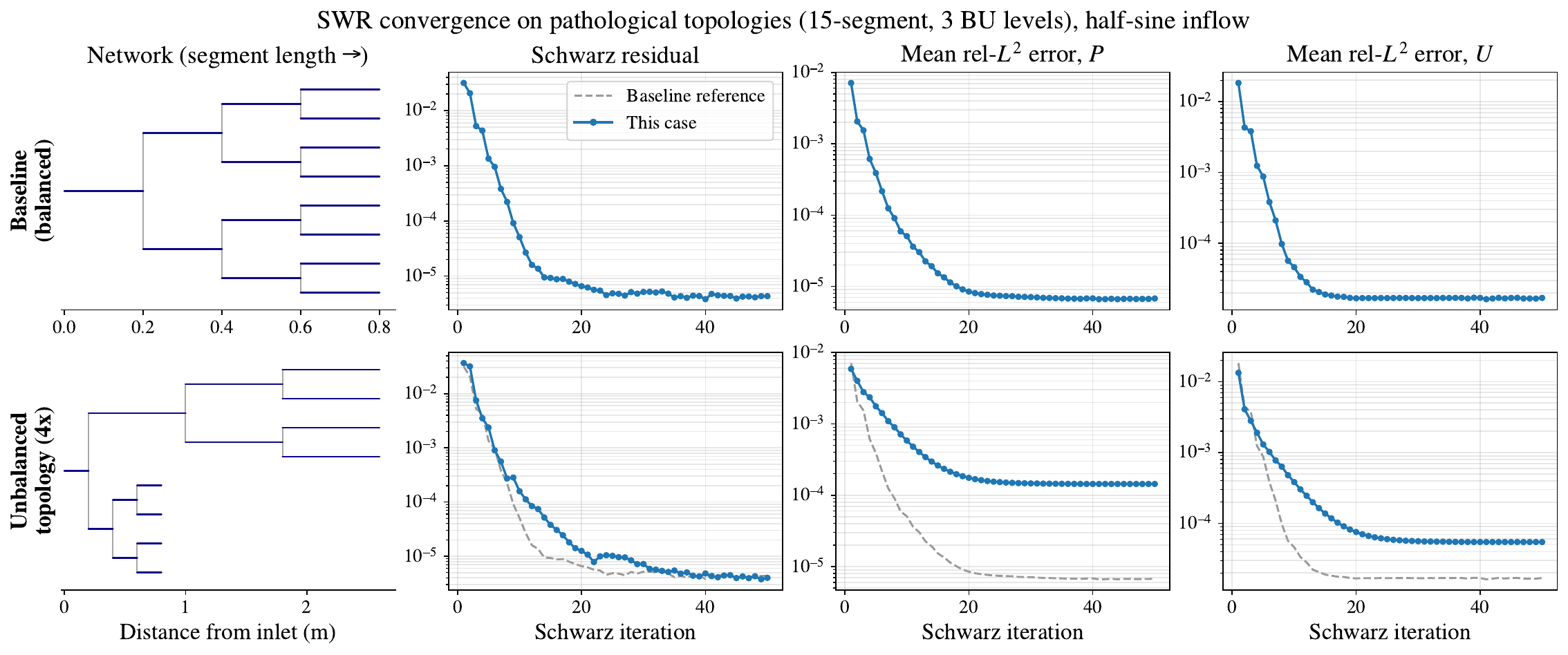} 
    \caption{SWR convergence on a balanced tree (top row) and a strongly unbalanced tree in which one first-generation subtree has segments four times longer than the other (bottom row), on the synthetic 15-segment tree. Left column: tree schematic. Remaining columns: self-convergence residual, mean relative $L^2$ error in $P$, and in $U$, versus Schwarz iteration.}
    \label{fig:swr_convergence_topology}
\end{figure}

For this and the following subsections of SWR convergence analysis, we use a balanced 15-segment synthetic tree (3 BU levels) with identical segments. The default configuration involves:
\begin{itemize}
    \item \textbf{Geometry:} segment length $L = 20$~cm, root lumen diameter $\approx 16$~mm, with the cross-sectional area halved at each bifurcation;
    \item \textbf{Wall:} purely elastic ($\gamma = 0$), with pulse-wave-velocity parameter $a = 11.0$;
    \item \textbf{Inlet:} a half-sine flow with cycle-mean $1\times10^{-4}$~m$^3$/s ($\approx 6$~L/min);
    \item \textbf{Terminal load:} a three-element Windkessel split in parallel across the leaves, with baseline totals $R_T \approx 1.1\times10^{8}$~Pa$\cdot$s$\cdot$m$^{-3}$ and $C_T \approx 1.3\times10^{-8}$~m$^3$/Pa.
\end{itemize}
Each run starts from the warm-start initialization, performs 50 Schwarz iterations, and simulates enough cardiac cycles for the peripheral flow to reach a periodic steady state. We report the self-convergence residual together with the mean relative $L^2$ error against a monolithic Nektar1D solve on the same tree. Each subsection then varies a single factor about this baseline while holding all others fixed.

Here we vary the tree structure, comparing the balanced baseline against an unbalanced tree in which one first-generation subtree has segments four times longer than the other. As observed in Figure~\ref{fig:swr_convergence_topology}, both cases converge, with the residual dropping to $\approx 4\times10^{-6}$. The balanced tree reaches a mean relative $L^2$ error near $7\times10^{-6}$, while the unbalanced tree settles higher at $\approx 1.4\times10^{-4}$ with its residual still dropping, showing that SWR converges even on a strongly unbalanced tree.

\subsection{Terminal resistance and compliance}
\label{app:swr_convergence_RC}

\begin{figure}[ht!]
    \centering
    \includegraphics[width=\linewidth]{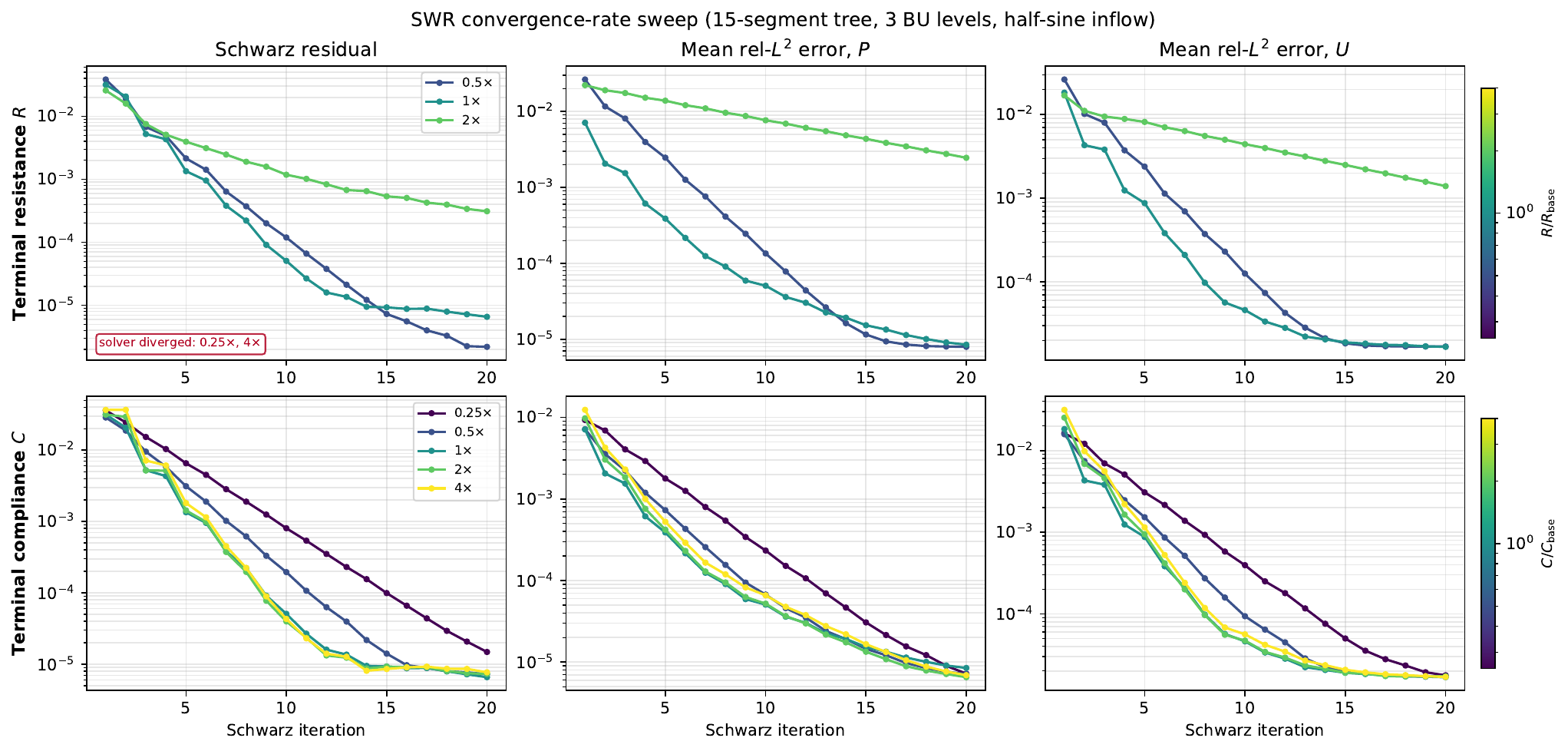}
    \caption{SWR convergence under a wide sweep of the terminal Windkessel
    resistance $R_T$ (top row) and compliance $C_T$ (bottom row), each scaled by
    $\{0.25, 0.5, 1, 2, 4\}$ about the baseline on a 15-segment synthetic tree
    (3 BU levels). Columns: self-convergence residual, mean relative $L^2$ error
    in $P$, and in $U$, versus Schwarz iteration.}
    \label{fig:swr_convergence_RC}
\end{figure}

Terminal resistance and compliance act on the simulation through different
mechanisms: $R_T$ sets the mean pressure ($\bar P \approx R_T \bar Q$),
while the product $\tau = R_T C_T$ sets the rate at which peripheral pressure decays between beats. Because this decay also governs the rate a simulation reaches a periodic steady state, we scale the total number of simulated cycles with $\tau$ so that the BU boundaries pass converged waveforms.

For the present subsection, we scale $R_T$ and $C_T$ independently by $\{0.25, 0.5, 1, 2, 4\}$ (one scalar per run), run 20 Schwarz iterations from the warm start, and report the residual and the mean relative $L^2$ error against a monolithic Nektar1D solve on the same tree.

As observed in Figure~\ref{fig:swr_convergence_RC}, compliance has almost no effect on convergence, with the mean relative $L^2$ error staying near $5\times10^{-6}$ across the full range. Resistance slows convergence slightly, with mid-range cases reaching $\approx 7\times10^{-6}$ and high resistance ($R_T{\times}2$) reaching $1.3\times10^{-4}$ by iteration~20 with the error still dropping. At the two resistance extremes, the system is not physiological, and the local numerical solver fails at the first iteration, before the Schwarz coupling starts.

\subsection{Vessel stiffness}
\label{app:swr_convergence_stiffness}

\begin{figure}[ht!]
    \centering
    \includegraphics[width=\linewidth]{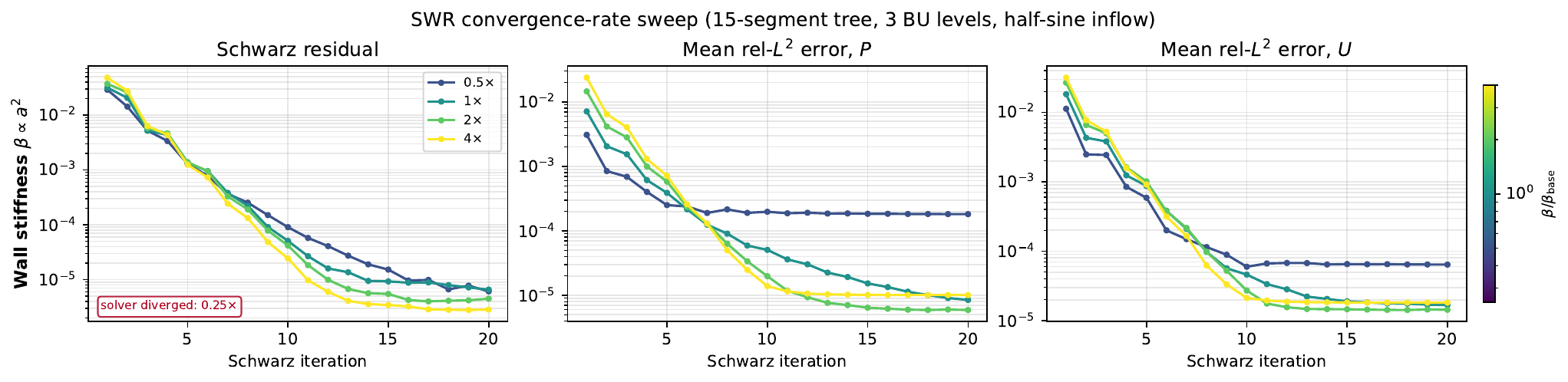}
    \caption{SWR convergence under a uniform scaling of the wall stiffness $\beta$ by $\{0.25, 0.5, 1, 2, 4\}$ on the synthetic 15-segment tree. Layout as in Figure~\ref{fig:swr_convergence_RC}.}
    \label{fig:swr_convergence_stiffness}
\end{figure}

Wall stiffness $\beta$ sets the pulse wave velocity ($c \propto \sqrt{\beta}$), which controls how strongly waves reflect at bifurcations. On the same synthetic tree (\ref{app:swr_convergence_RC}) we scale $\beta$ by $\{0.25, 0.5, 1, 2, 4\}$, changing each element's wave speed while leaving the total $R_T$, $C_T$ unchanged. As observed in Figure~\ref{fig:swr_convergence_stiffness}, SWR converges across the range, with the baseline and stiffer walls reaching $\approx 5\text{--}10\times10^{-6}$ and the soft wall ($\beta{\times}0.5$) reaching $\approx 1.8\times10^{-4}$ while its residual keeps dropping. At the softest value ($\beta{\times}0.25$) the monolithic reference solver crashes at the first time step, so no reference exists to compare against, which is a limit of the numerical solver at a very soft wall, not of the SWR coupling.

\subsection{Heart rate}
\label{app:swr_convergence_heartrate}

\begin{figure}[ht!]
    \centering
    \includegraphics[width=\linewidth]{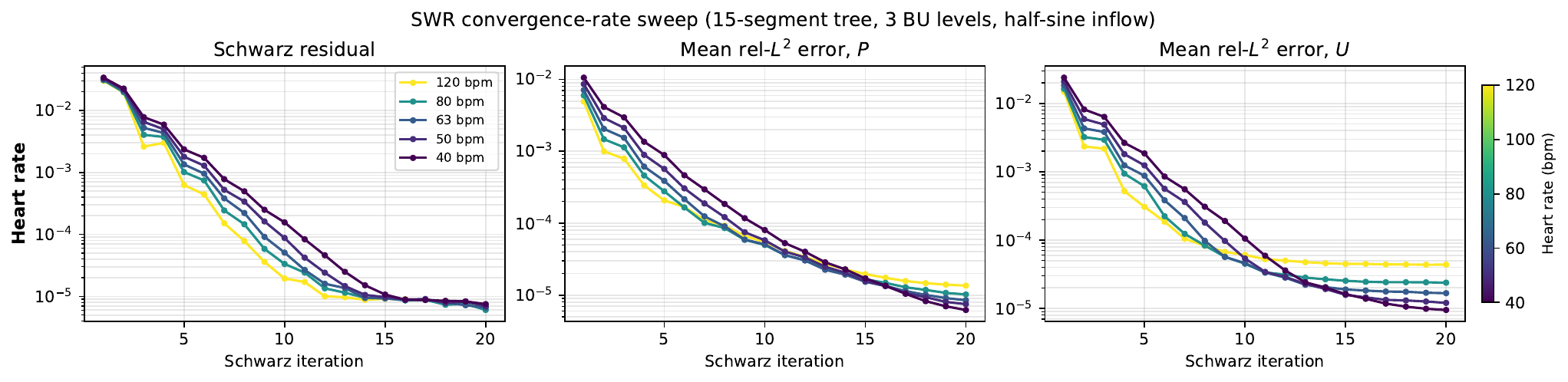}
    \caption{SWR convergence across cardiac periods $T_{\mathrm{cardiac}} = 1.5$--$0.5$~s (40--120~bpm) on the synthetic 15-segment tree. Layout as in Figure~\ref{fig:swr_convergence_RC}.}
    \label{fig:swr_convergence_heartrate}
\end{figure}

The heart rate sets the cardiac period $T_{\mathrm{cardiac}}$, which is also the SWR exchange window $[0, T_{\mathrm{cardiac}}]$ over which the waveforms are relaxed, so this sweep additionally probes sensitivity to the length of the coupling window. On the same synthetic tree we vary $T_{\mathrm{cardiac}}$ from $1.5$ to $0.5$~s (40 to 120~bpm), with the half-sine inlet period tracking $T_{\mathrm{cardiac}}$. As observed in Figure~\ref{fig:swr_convergence_heartrate}, multiplicative SWR converges cleanly at every rate, reaching a mean relative $L^2$ error of $\approx 3\text{--}12\times10^{-6}$ with no diverged cases.

\subsection{Inflow waveform}
\label{app:swr_convergence_inflow}

\begin{figure}[ht!]
    \centering
    \includegraphics[width=\linewidth]{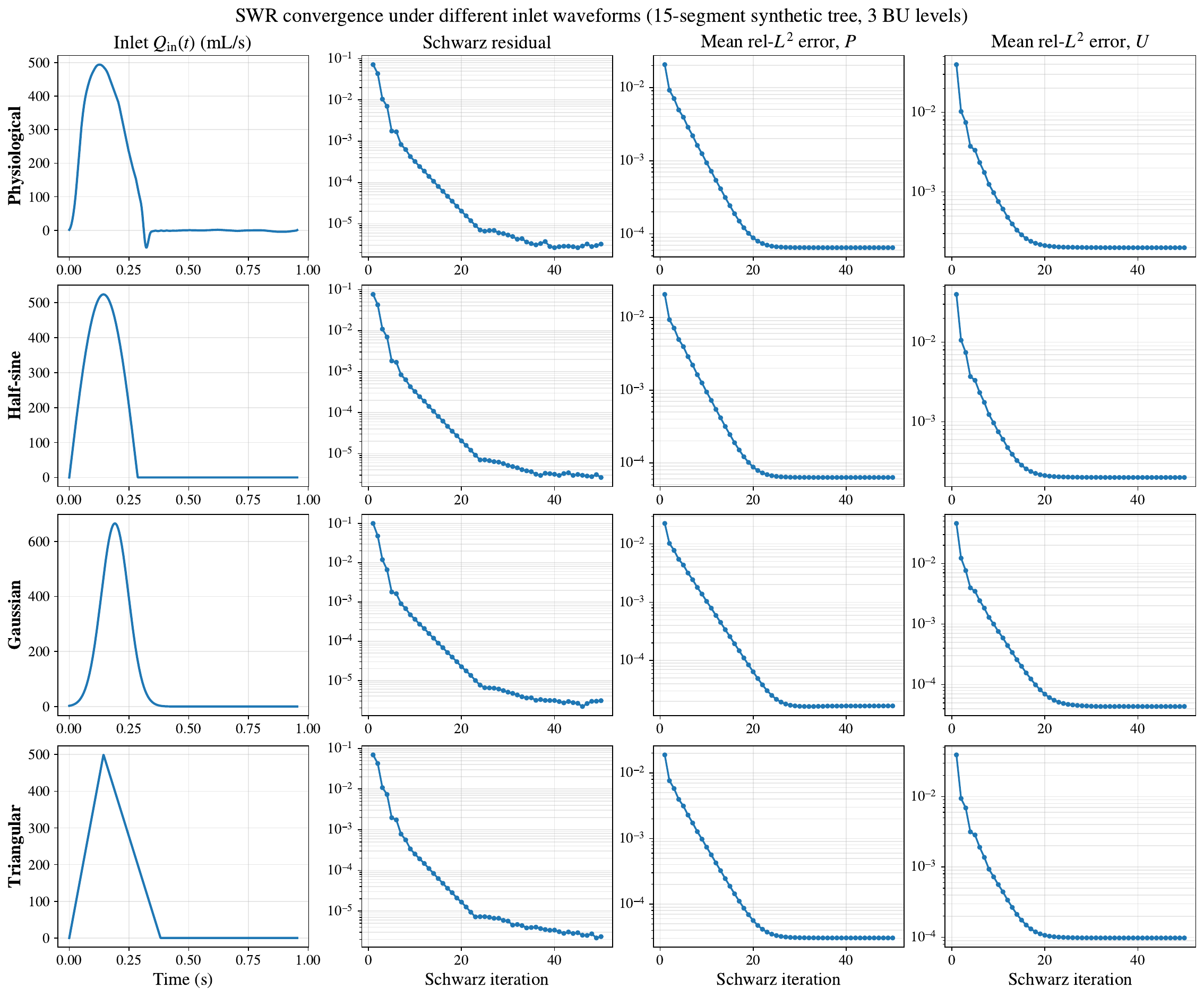}
    \caption{SWR convergence under four structurally distinct root inflow waveforms $Q_{\mathrm{in}}(t)$ --- the $40$-harmonic physiological waveform (control), a half-sine systolic ejection, a narrow Gaussian pulse, and a triangular pulse, all rescaled to matched cycle-mean flow --- on the synthetic 15-segment tree, one shape per row. Left column: the inlet waveform. Remaining columns: self-convergence residual, mean relative $L^2$ error in $P$, and in $U$, versus Schwarz iteration.}
    \label{fig:swr_convergence_inflow}
\end{figure}

Finally, we vary the root inflow waveform while holding the network and all physical parameters fixed, to confirm that convergence does not depend on the specific driving signal. On the same synthetic 15-segment tree we replace the inlet $Q(t)$ with four structurally distinct shapes: the $40$-harmonic physiological waveform used for our dataset, a half-sine systolic ejection, a narrow Gaussian pulse, and a triangular pulse. Each waveform is rescaled to the same cycle-mean flow, so the comparison isolates pulse shape from stroke volume and heart rate. As observed in Figure~\ref{fig:swr_convergence_inflow}, all four shapes converge, with the residual dropping to $\approx 2\text{--}3\times10^{-6}$ in every case. The mean relative $L^2$ pressure error settles between $\approx 1.7\times10^{-5}$ and $\approx 6.5\times10^{-5}$ depending on the shape, all well below $10^{-4}$, confirming that convergence is robust to the inflow waveform.

\section{Parallelized Throughput and Hardware-Usage Disclosure}
\label{app:parallel_timing}

All SWR operator timings in this work are measured on a single NVIDIA A100 (40\,GB), with auxiliary CPU threads used only for data I/O and host-device transfer. The reference Nektar1D timings are measured on a single CPU core, with one independent simulation running per core. Throughput on a multi-core CPU therefore scales proportionally with the core count when independent cases are processed in parallel. The reported speedups should be read as a deployment-level comparison between the operator on a GPU and the numerical solver on a CPU core, not as a per-FLOP efficiency measurement.

Because the per-case operator forward pass uses only a small fraction of the A100 memory budget, many cases can be evaluated concurrently on the same device. Table~\ref{tab:batched_per_case} reports the SWR operator wall-clock time on the 55-segment model at three batch sizes $B$ in float32 (fp32) and at $B=128$ in bfloat16 (bf16), using the 30 Schwarz iterations adopted in Section~\ref{sec:network_performance}. Timings exclude a small per-case data-loading overhead that is not part of the solver cost. In fp32, the per-case time drops from $6.30$\,s at $B=1$ to $2.05$\,s at $B=128$, reflecting the saturation of the A100 beyond $B\!\approx\!32$. Switching inference to bf16 at $B=128$ further reduces the per-case time to $0.54$\,s, a $3.8\times$ speedup over the fp32 baseline with only a $\sim\!0.2$ percentage-point increase in the relative $L^2$ errors for both pressure and velocity.

At this throughput, the batched operator matches the amortized cost of running Nektar1D concurrently on roughly $113.30 / 0.54 \approx 210$ CPU cores, that is, a modest HPC allocation. An HPC node of this size is typically $3\text{--}4\times$ more expensive than a single A100 (40\,GB) in both capital and operational cost, so the reported speedup translates into a substantial hardware-cost advantage at equivalent throughput, in addition to the wall-clock improvement.

\begin{table}[ht!]
\centering
\footnotesize
\caption{Wall-clock time and accuracy of the SWR operator solver on the 55-segment model for 30 Schwarz iterations, measured on one NVIDIA A100 (40\,GB). Per-case time is total time divided by $B$. Relative $L^2$ errors in pressure ($P$) and velocity ($U$) are averaged over all test cases.}
\label{tab:batched_per_case}
\begin{tabular}{cccccc}
\toprule
dtype & $B$ & Wall-clock (s) & Per-case (s) & Rel.\ $L^2$ $P$ (\%) & Rel.\ $L^2$ $U$ (\%) \\
\midrule
\multirow{3}{*}{fp32} & 1   & 6.30   & 6.30 & \multirow{3}{*}{\textbf{1.44}} & \multirow{3}{*}{\textbf{2.29}} \\
                      & 32  & 67.77  & 2.12 & & \\
                      & 128 & 262.92 & 2.05 & & \\
\midrule
bf16                  & 128 & \textbf{69.30}  & \textbf{0.54} & 1.60 & 2.49 \\
\bottomrule
\end{tabular}
\end{table}

\section{Ablation Studies on Backbone}
\label{app:backbone}

The proposed BU-based framework is independent of the specific neural operator backbone. Although we employ a transformer-based operator as the main implementation due to its strong representation capability, the proposed decomposition and Schwarz coupling strategy can also be combined with alternative operator-learning architectures. Here, we replace BUFormer with a standard Fourier Neural Operator (FNO)~\cite{li2020fourier}, while retaining the bifurcation-unit decomposition and Schwarz waveform relaxation coupling. Because a plain FNO requires a fixed input/output structure and cannot accommodate the BC-dependent boundary layouts handled by BUFormer's masked encoder, we train a separate FNO for each BC type (\emph{FNO-per-type}), with the number of trainable parameters matched to the corresponding BUFormer for each tree. The training protocol is unchanged: relative-$L^2$ loss, AdamW with cosine scheduling, and BC-type-balanced sampling.

\begin{table}[h!]
\centering
\footnotesize
\setlength{\tabcolsep}{4pt}
\caption{Full-network SWR accuracy of the FNO-per-type backbone on the 7-segment and 55-segment trees, with the default BUFormer for reference. Mean relative $L^2$ error over the test cases (in $10^{-2}$), at 20 (7-segment) and 30 (55-segment) Schwarz iterations.}
\label{tab:backbone}
\begin{tabular}{lcccc}
\toprule
& \multicolumn{2}{c}{7-segment} & \multicolumn{2}{c}{55-segment}\\
\cmidrule(lr){2-3}\cmidrule(lr){4-5}
Backbone & $P$ & $U$ & $P$ & $U$\\
\midrule
FNO-per-type       & 2.35 & 2.65 & 3.64 & 2.49\\
BUFormer (default) & 0.32 & 0.38 & 1.15 & 2.21\\
\bottomrule
\end{tabular}
\end{table}

Table~\ref{tab:backbone} reports the full-network SWR accuracy of the FNO-per-type backbone on both trees. The bifurcation-unit decomposition and Schwarz coupling remain functional with this alternative backbone, confirming that the framework is not tied to a specific operator: any operator that fits the local BU physics can be embedded in the SWR iteration to produce a network-level solution. We adopt BUFormer as the default backbone because its shared, mask-based encoder handles all boundary-condition combinations within a single model, whereas FNO-per-type requires a separate model for each BC type.

\section{Ablation Studies on Fourier Mode and Balanced Sampling}
\label{app:ablation}

We ablate two design choices of the BUFormer operator: the number of retained Fourier modes $K$ in the spectral encoder/decoder (\eqref{eq:fourier_enc}--\eqref{eq:fourier_dec}), and the balanced sampling over boundary-condition (BC) types (Section~\ref{sec:experiments}). Both are studied in the lighter 7-segment setting, varying one knob at a time with all other hyperparameters fixed. We report the relative $L^2$ error on individual BUs, either per field (pressure $P$, velocity $U$) or as the full state $\mathbf{S}$.

\subsection{Fourier modes}
\label{app:ablation_fourier}

\begin{figure}[t!]
    \centering
    \includegraphics[width=0.75\linewidth]{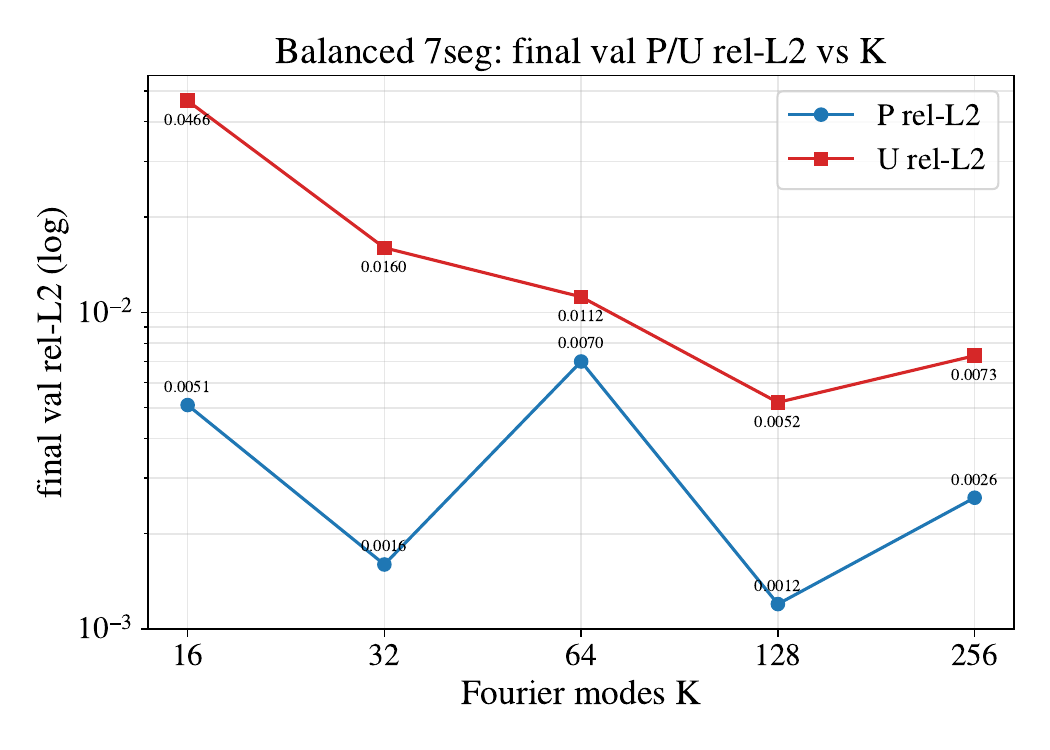}
    \caption{Final validation relative $L^2$ error for pressure $P$ and velocity $U$ versus the number of retained Fourier modes $K$ (7-segment model, balanced sampling). Error decreases with $K$ up to $K=128$ and saturates thereafter.}
    \label{fig:ablation_ksweep}
\end{figure}

Figure~\ref{fig:ablation_ksweep} sweeps $K \in \{16,32,64,128,256\}$. The dominant velocity error decreases monotonically from $4.7\times10^{-2}$ ($K=16$) to $5.2\times10^{-3}$ ($K=128$) and reverses at $K\!=\!256$, confirming that the low-frequency modes carry essentially all of the quasi-periodic signal while higher modes fit noise. The pressure error stays below $1\times10^{-2}$ and is likewise smallest at $K=128$. We adopt $K=128$ for the main model (\ref{app:hyperparams}).

\subsection{Balanced sampling over BC types}
\label{app:ablation_balance}

\begin{figure}[t!]
    \centering
    \includegraphics[width=0.75\linewidth]{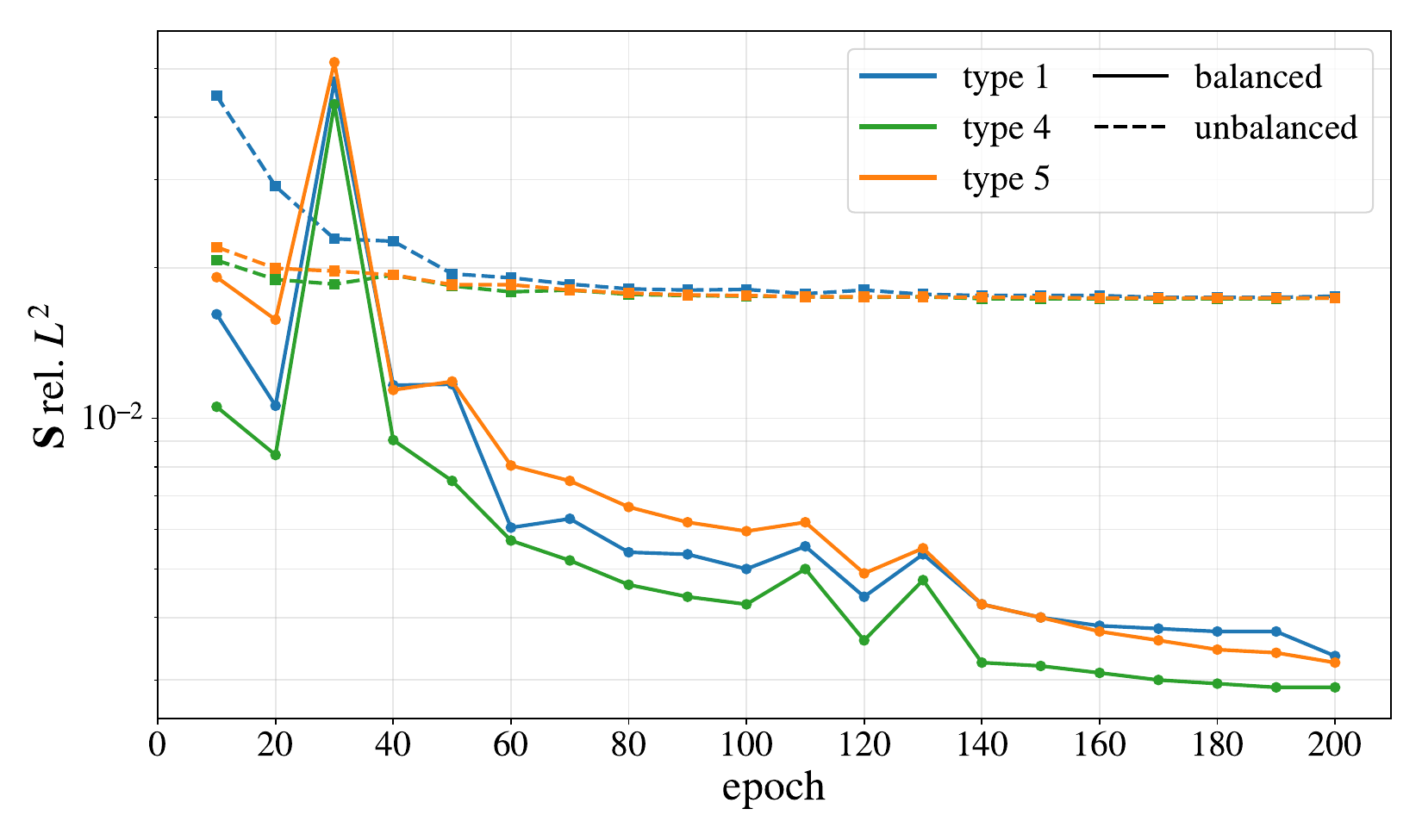}
    \caption{Per-type full state $\mathbf{S}$ relative $L^2$ error versus epoch, with (solid) and without (dashed) balanced sampling, for the three BC types present in the 7-segment tree (Table~\ref{tab:bu_types}). Without balancing, all types stall; with balancing, every type keeps improving by roughly an order of magnitude.}
    \label{fig:ablation_balance}
\end{figure}

As noted in Section~\ref{sec:experiments}, the SWR algorithm produces a strongly imbalanced mixture of BC types (Table~\ref{tab:bu_types}): rare but structurally critical configurations such as the root type~1, which initiates the warm-start, are heavily under-represented. Figure~\ref{fig:ablation_balance} tracks the per-type field-averaged error over training with and without balanced sampling. Without balancing, all three types stall at $\approx 1.8\times10^{-2}$ after $\sim\!60$ epochs, with the rare root type~1 the worst of the three; with balancing, every type continues to improve to $\approx 2\text{--}3\times10^{-3}$, roughly an order of magnitude lower. Balanced sampling therefore not only protects the rare root configuration but improves all types, confirming it as a necessary ingredient rather than a minor tweak.

\end{document}